\documentclass[reprint,secnumarabic,amssymb, nobibnotes, aps, prb]{revtex4-1}

\usepackage{graphicx}
\usepackage{graphics}
\usepackage[percent]{overpic}
\graphicspath{ {images/} }
\usepackage{amsmath}
\usepackage{mathtools}
\usepackage[utf8]{inputenc}
\usepackage{amsfonts}
\usepackage{amssymb}
\usepackage{float}
\usepackage{comment}
\usepackage{color,soul}
\usepackage{textcomp}
\usepackage{subcaption}
\usepackage{hyperref}
\usepackage[dvipsnames]{xcolor}
\newcommand{\vM}{{\bf M}}
\newcommand{\vQ}{{\bf Q}}
\newcommand{\vk}{{\bf k}}
\newcommand{\vp}{{\bf p}}
\newcommand{\vm}{{\bf m}}
\renewcommand{\vr}{{\bf r}}
\newcommand{\cH}{\mathcal{H}}

\newcommand{\cC}{\mathcal{C}}

\newcommand{\cTg}{\mathcal{T}_\pi}
\newcommand{\vsigma}{\mbox{\boldmath$\sigma$}}
\newcommand{\grad}{\mbox{\boldmath$\nabla$}}
\newcommand{\id}{\,\mathrm{d}}
\newcommand{\intk}{\,\int  \frac{\id ^2 k}{8\pi^2}}

\newcommand{\intkprfbz}{\,\int\frac{\id ^2 k'}{(2\pi)^2}}

\DeclareMathOperator{\sgn}{sgn}

\begin{document}

\title{Thermal Transport in Superconductors with coexisting Spin Density Wave Order}%

\author{Sourav Sen Choudhury, Anton B. Vorontsov}%
\affiliation{Department of Physics, Montana State University, Bozeman, Montana 59717, USA}
\date{\today}
\begin{abstract}
        We study thermal transport in a two-dimensional system with coexisting $s$- or $d$-wave Superconducting (SC) and Spin Density Wave (SDW) orders. We analyse the nature of coexistence phase in a tight-binding square lattice with $\vQ=(\pi,\pi)$ SDW ordering. The electronic thermal conductivity is computed within the framework of the Boltzmann kinetic theory, using Born approximation for the impurity scattering collision integral. We describe the influence of the Fermi surface (FS) topology, the competition between the SC and SDW order parameters, the presence or absence of zero energy excitations in the coexistence phase, on the low temperature behavior of thermal conductivity of the various paring states. We present qualitative analytical, and fully numerical results that show that the heat transport signatures of various SC states emerging from collinear SDW order are quite distinct, and depend on the symmetry properties of the SC order parameter under translation by the SDW nesting vector $\vQ$. {A combination of $(\pi,\pi)$-SDW and the $d_{x^2-y^2}$ pairing state results in fully gapped excitations, whereas $(\pi,\pi)$-SDW co-existing with either $d_{xy}$ or $s$-wave pairing states may always have gapless excitations. There appear special stable Dirac nodal points that are not gapped by the SC order in the coexistence phase, resulting in finite residual heat conductivity. 
        } 
\end{abstract}
\maketitle

\section{Introduction}

In normal metals, at low temperatures transport properties are primarily determined by scattering of electrons by impurities. The thermal conductivity $\kappa(T)$ has a linear $T$ dependence, which is well understood within the framework of semi-classical transport theory based on the Boltzmann kinetic equation\cite{ziman}. The kinetic formulation was also successfully used to explain the effects that conventional superconductivity has on the thermal conductivity\cite{bardeen,gel}. With the discoveries of heavy fermion \cite{hfrev}, cuprate  \cite{harlin, tsu, tail, pair} and iron based \cite{Wen, stewart, chubukov} superconductors, new questions have arisen with regards to the low temperature transport properties of superconductors. 
The behavior of the thermal conductivity at low temperatures for these unconventional superconductors is not at all like that of the fully-gapped conventional type superconductors. 
One reason is that most unconventional superconductors have a nodal gap structure i.e there exist points on the Fermi surface (FS), nodes, where the superconducting gap is zero. As the energy gap is small around the nodes, the nodal quasiparticles can be easily excited and they dominate the heat transport properties of such superconductors. This problem has been studied by a number of authors at various levels of complexity \cite{arfi, Hirsch,scharn,mon, durst, graf}, and thermal conductivity measurements became a very useful probe of superconductivity as it can reveal the gap structure of unconventional superconductors\cite{Matsuda, Shak}. 

Another characteristic feature of many unconventional superconductors is the proximity of magnetic and superconducting orders in these materials \cite{lake,mathur,Badoux2016,kim,ley}. 
The electronic phase diagrams of many highly correlated systems are complex, with multiple broken symmetry phases appearing with similar 
ordering temperatures as material properties, such as dopant concentration, are varied over wide ranges. For example, there is a proximate antiferromagnetic (AF) state in the phase diagrams of superconductors such as cuprates \cite{Badoux2016}, iron pnictides \cite{ley} and heavy fermion superconductors\cite{mathur, kim}. 


{However much less is known about  thermal transport in superconductors with coexisting orders (Spin-Density wave - SDW, Charge-density wave - CDW). 
Previous studies have addressed mainly one aspect of the heat conductivity in coexisting phases like superconducting (SC) and CDW or SC and SDW \cite{durst2, Schiff}, with cuprates as an application. In $d$-wave superconductors heat transport by nodal quasiparticles shows impurity-independent, \emph{universal}, limit at low temperatures \cite{durst, graf}, seen in many materials \cite{sun,Shak}. 
Theoretical investigations of thermal transport in ``superconductor + density wave order'' systems \cite{durst2, Schiff} attacked the issue of CDW or SDW order influencing the $T\to 0$ limit of thermal conductivity by nodal quasiparticles in $d_{x^2-y^2}$ superconductors, in particular how the nodes get gapped by the additional order. 
The transport calculations were carried out in $2D$, within Kubo linear response theory using Green's function technique, where impurity effects were included only through non-self-consistent energy broadening parameter. The CDW or SDW were also incorporated non-self-consistently, as an additional tunable small order on top of the SC state, and neither the nature of the co-existence, nor its temperature dependence, was investigated. 
These calculations indicated that the robustness of the universal limit of thermal conductivity of $d_{x^2-y^2}$ superconductors depends on direction of the ordering vector, and the type of the coexisting order (CDW or SDW). 
Additional order displaces the nodes in $k$-space. For example, the nodal quasiparticles become gaped by SDW once the $d$-SC nodes are separated by exactly the ordering $\vQ$ vector \cite{Schiff}.
Another study\cite{Chatterjee} looked at the changes in zero-temperature heat transport across continuous SC to SC+SDW transition for $d_{x^2-y^2}$ superconductor, employing the same non-self-consistent treatments of impurities and SDW order, assumed to be controlled by doping. 
These calculations show that thermal conductivity behaves very differently depending whether emerging SDW is commensurate or incommensurate. For a commensurate SDW the SC $\to$ SC+SDW transition results in a gradual drop in $\kappa$ as a function of growing SDW order, whereas incommensurate SDW results in a sharp drop across the transition.\cite{Chatterjee} 
} 

In this paper we look at the thermal transport properties of a number of different superconducting states in which the SC order coexists with the anti-ferromagnetic spin density wave (SDW) order in the full temperature range. 
For transport calculation we use quasiparticle Boltzmann equation which is physically more transparent than the Green's function or quasi-classical techniques. {The goal is to understand the nature of the  different coexistence states arising from the interplay between the SC and SDW order parameters, and its impact} on the temperature behavior of the thermal conductivity of several paring states: $s$-wave, $d_{x^2-y^2}$ and $d_{xy}$ symmetry. The choice of the $d$-wave states is motivated by the fact that it is a prototypical unconventional pairing state, with sign-changing order parameter and nodal quasiparticles, applicable to heavy fermion and cuprate superconductors\cite{tsu}. In this paper we calculate the thermal conductivity in which scattering of quasiparticles by nonmagnetic impurities is the dominant process. Within Boltzmann theory, we only consider the case of small phase shifts i.e. the Born approximation for weak interaction of electrons with impurities. 
The impurity scattering is the dominant mechanism at lower temperatures, and at higher temperatures the scattering rates can be augmented to incorporate temperature-dependent inelastic scattering\cite{Graf1996, Graf2000}, to reflect the growth of thermal conductivity below $T_c$, seen in e.g. ${\mathrm{CeCoIn}}_{5}$ \cite{kim}, or UPt$_3$ \cite{Luss}.  
%

The organization of the paper is as follows. In sections \ref{sec:model}.\ref{sec:H}-\ref{sec:FS}, we discuss the model Hamiltonian,  symmetries of the SC order parameter and the topology of the Fermi surface. 
Self-consistent approach to determining co-existing SDW and SC order parameters is presented in section \ref{sec:model}.\ref{sec:selfconsistency}. 
Kinetic formalism  
is described in section \ref{sec:model}.\ref{sec:kinetic}. 
Numerical results for heat conductivity is discussed in section \ref{sec:results}. 
Section \ref{sec:conclusion} is a brief conclusion.

\section{Model and Formalism}
\label{sec:model}

\subsection{\label{sec:H} Hamiltonian}

For our model we start with a tight-binding normal state Hamiltonian 
\begin{equation}
 H_0 =\sum\limits_{\vk ,\sigma = \pm1} \xi({\vk })c_{\vk \sigma}^{\dagger}c_{\vk \sigma},   
\end{equation}
where 
$$ 
\xi(\vk )= -t_1(\cos k_x+ \cos k_y) -t_2\cos k_x \cos k_y - \mu 
$$ 
is the inversion-symmetric dispersion relation, $\xi(\vk )= \xi(-\vk )$. 
It describes the nearest neighbour ($t_1>0$) and next-nearest neighbour ($t_2>0$) hopping on a 2D square lattice with lattice spacing $a=1$. We set the chemical potential to zero, and therefore for $t_2>0$ the electron filling is slightly less than half. 
This results in a Fermi surface that is not perfectly nested and therefore potentially susceptible to coexistence of SC and SDW order parameters. The perfect nesting limit is given by setting $t_2=0$. The coexistence of SC and SDW orders in models of this type have been previously studied by Machida\cite {machida, machida2,machida3}. 
We wish to look at heat transport in these models across the SDW $\to$ SC transition. The full mean-field Hamiltonian for a system with intertwined SC and SDW order is given by \cite{machida3}
\begin{eqnarray}
\begin{aligned}
&H = H_0 + H_{SDW} +H_{SC},\\ 
&H_{SDW} = \frac12 \sum\limits_{\vk ,\sigma }\sigma M \left( c_{\vk \sigma}^{\dagger}c_{\mathbf{k +Q}\sigma} + h.c. \right),\\ 
&H_{SC} = \frac12 \sum\limits_{\vk ,\sigma} \sigma \Delta_{\vk} \left(c_{\vk \sigma}^{\dagger}c_{-\vk  -\sigma}^{\dagger} + h.c.\right). 
\label{ham full}
\end{aligned} 
\end{eqnarray}
The mean field order parameters are defined by the following self consistent equations 
\begin{align}
\begin{split}
M &=-\frac{U}{2}\sum\limits_{\vk ,\sigma } \sigma\langle c_{\mathbf{k+Q}\sigma}^{\dagger}c_{\vk \sigma}\rangle, \\
\Delta_{\vk}&=-\sum\limits_{\vk' }g(\mathbf{k,k'})\langle c^{\dagger}_{\mathbf{-k'},\downarrow}c^{\dagger}_{\vk' ,\uparrow}\rangle \\
\end{split}
\label{M&delta}
\end{align}
where $U$ is the repulsive on-site Coulomb interaction which leads to the SDW formation.  We consider a collinear sinusoidal SDW with spatial magnetization $\mathbf{m(r)}= 2M\mathbf{\hat{z}}\, \cos(\mathbf{Q \cdot r})$. The SDW couples electron states with parallel spins and momenta differing by the nesting vector $\mathbf{Q}$, i.e. $(\vk \uparrow)$ with $(\mathbf{k+Q}\uparrow)$ and  $(\vk \downarrow)$ with $(\mathbf{k+Q}\downarrow)$ (this is schematically represented by dashed lines in Fig.~\ref{FS}). As for the SC pairing interaction, we consider the singlet channel and assume the interaction to be of the form $g(\mathbf{k,k'})=g\eta(\vk ) \eta(\vk' )$, $\eta(\vk )$ being a basis function compatible with the square symmetry of the 2D lattice. The SC order parameter combines time-reversed electron states with opposite momenta and anti-parallel spins, i.e. $(-\vk \downarrow)$ with $(\vk \uparrow)$ and $(\mathbf{-k-Q}\downarrow)$ with $(\mathbf{k+Q}\uparrow)$ (this is schematically represented by dotted lines in  Fig.~\ref{FS}). For our purposes we consider the case where the pure SDW transition temperature $T_{SDW}$ is greater than the pure SC transition temperature  $T_{C0}$, i.e the ratio 
$$p=\frac{T_{C0}}{T_{SDW}}<1 \,.$$

\subsection{Symmetry Classes of the SC Order Parameters}
\label{sec:OPclass}

The coexistence problem  critically depends on the symmetry properties of the SC order parameter and also on the topology of the FS. If we consider the case of a commensurate SDW with  nesting vector $\mathbf{Q}=(\pi,\pi)$ i.e $2\mathbf{Q}= \mathbf{G} = (2\pi, 2\pi)$ - the diagonal reciprocal lattice vectors for the 2D square lattice, - then the various SC paring states can be classified \cite{Kato} based on the combined symmetry operations of parity: $\Delta_{-\vk}=\pm\Delta_{\vk} $ (even or odd) and translation by the nesting  vector: $\Delta_{\mathbf{k+Q}}=\pm\Delta_{\vk} $ (even or odd). The symmetry classification of the paring states ($\Delta_{\vk} =\Delta \eta(\vk )$) are summarized in Table \ref{tab1}. This classification  has important consequences for the coexistence problem \cite{machida3}: the SC states in the \textit{(E, E)} class are competitive with the SDW, whereas states in the \textit{(E, O)} class are less competitive with the SDW and the two orders can naturally co-exist. The difference in the nature  of the coexistence problem in these two  distinct symmetry classes  has an obvious impact on the thermal transport properties of the system across the SDW\textrightarrow SC  transition. One of the aims of this paper is to establish the relation between the nature of the SC-SDW coexistence and its signatures in the electronic thermal transport.  
\begin{table}[t]
\begin{center}
\bgroup
\def\arraystretch{1.5}
\begin{tabular}{|p{2cm}|p{2cm}|p{4cm}|}
\hline
Symmetry Class & Pairing \newline State & SC Basis Function \\
\hline 
\textit{I or (E, O)} &  $d_{x^2-y^2}$ & $\eta(\vk )= \frac{1}{2}(\cos k_x - \cos k_y$)\\ 
\hline
\textit{II or (E, E)} & $d_{xy}$ \newline  $s$-wave  &$\eta(\vk )= \sin k_x \sin k_y$ \newline $\eta(\vk )=1$ \\ 
\hline
\end{tabular}
\egroup
\end{center}
\caption{The symmetry classification of the various paring states and the corresponding basis functions. First letter (E-even, O-odd) corresponds to parity symmetry $\Delta_{-\vk}=\pm\Delta_{\vk }$, and the second letter is for SDW translations $\Delta_{\mathbf{k+Q}}=\pm\Delta_{\vk }$. }
\label{tab1}
\end{table}
\subsection{Topology of the Fermi Surface}
\label{sec:FS}

\begin{figure}
\begin{center}
\includegraphics[width=\linewidth]{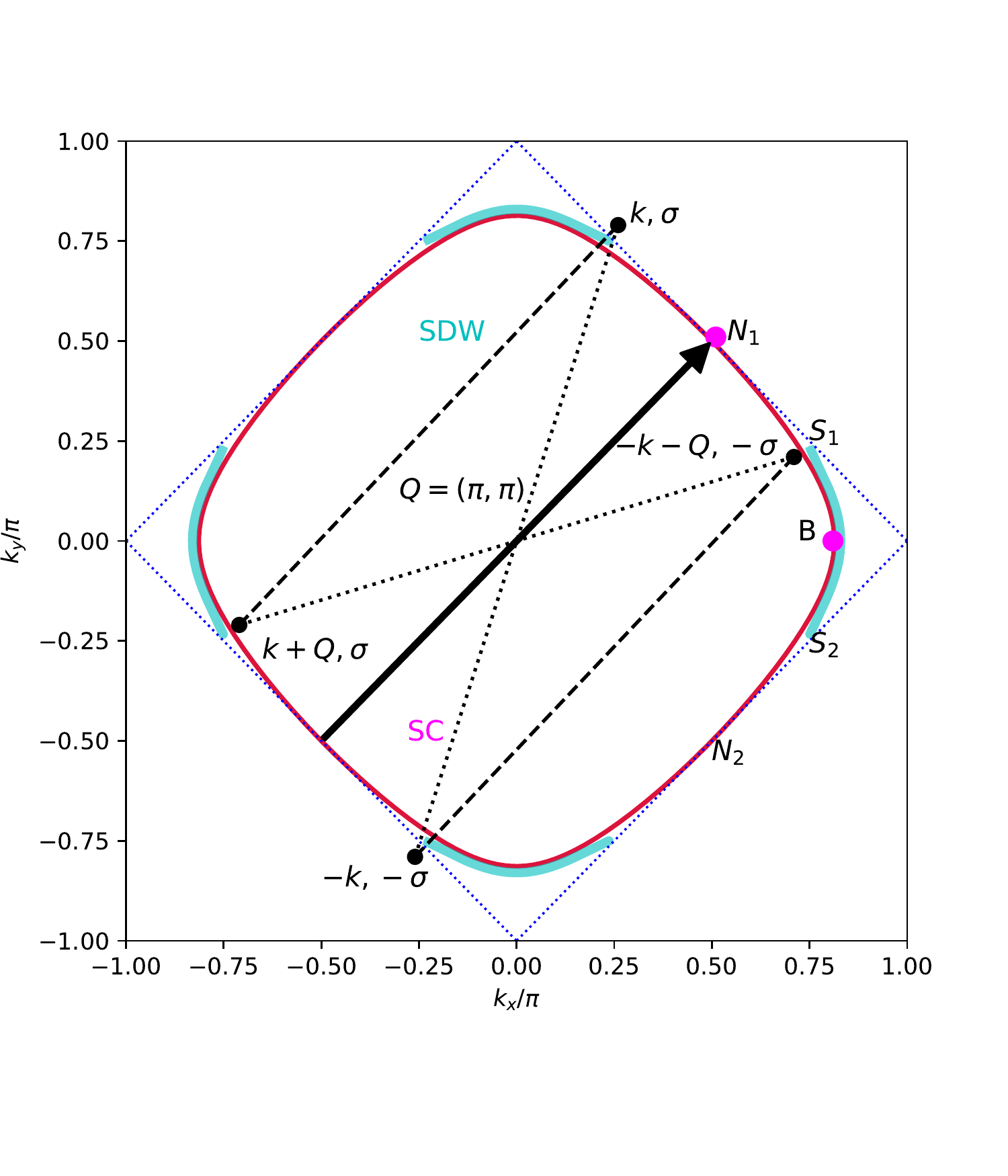}
\vspace{-1.5cm}
\caption{The FS of the normal state (solid red curve) and in the SDW phase (cyan curve). The parameters are $t_2/t_1=0.2$, $M/t_1=0.1$ (this value is taken for illustration purposes, the typical value in the calculations are $M/t_1 \sim 10^{-3}$). $\mathbf{Q}= (\pi,\pi)$ is the nesting vector. The dashed blue square indicates the boundary of the reduced Brillouin zone (RBZ).
}
\label{FS}
\end{center}
\end{figure}

In the Fig.~\ref{FS} we show the Fermi surface for our model. In the normal state the FS is indicated by the solid red curve. The SDW with the ordering vector $\mathbf{Q}=(\pi,\pi)$ doubles the lattice cell size reducing the Brillouin zone to the dotted blue square (RBZ). For this SDW ordering all 4 flat sides of the normal FS are nested, and become gapped, leaving zero energy excitations only at the corners -- the FS in the SDW state is indicated by the solid cyan curve. As the SDW gap grows from zero and reaches its maximum value the FS continuously shrinks from the $N_1-N_2$ section to the $S_1-S_2$ section. Points $N_1$ and $B$  denote location of the nodes of the $d_{x^2-y^2}$ and $d_{xy}$ paring states respectively, they are indicated by the magenta dots. Since SDW gaps region around point $N_1$, appearance of the $d_{x^2-y^2}$ SC gap completely removes the low-energy excitations. 
In the case of the $d_{xy}$ pairing state the low-energy excitations remain since the nodal line crosses Fermi pocket at point $B$, 
which is not gapped by the SDW. 
{Further, in the case of the $d_{xy}$ and $s$-wave pairing states, we show that unusual zero energy excitations remain on the boundary of the RBZ near points $S_{1,2}$, even when the SC order starts to grow inside the SDW state. The stability of these zero energy excitations is related to the even symmetry of the $d_{xy}$ and $s$-wave states under translations by the nesting vector $\vQ$. We explain this in more detail the following section. 
The relative positions of the nodes for the two $d$-wave pairing states, and the extra zero energy exciations  in the coexistence phase of the $d_{xy}$ state leads to their different thermal conductivity $\kappa(T)$ behavior.
}

\subsection{Diagonalization of the Model Hamiltonian} 
\label{sec:diagH}

The more general form for Hamiltonian (\ref{ham full}), corresponding to SDW magnetization $\mathbf m(\mathbf r) = Re(\vM_\vQ e^{i\vQ \mathbf r})$ is a $8\times8$ matrix 

\begin{align*}
H=&\frac{1}{4}\sum\limits_{\vk \in FBZ} \Psi^{\dagger}_{\vk } \cH_{\vk,\vQ } \Psi_{\vk }
\end{align*}
\vspace{-.5cm}
\begin{small}
\begin{align}
\cH_{\vk,\vQ }=&
\left( \begin{array}{c|c}
\begin{array}{cc}
\xi_{\vk } & \Delta_{\vk } (i\sigma_y) \\ 
-\Delta^*_{-\vk }(i\sigma_y) & -\xi_{-\vk }
\end{array}
&
\begin{array}{cc}
\vM_\vQ ^* \vsigma & 0  \\
0 & -\vM_\vQ^* \vsigma^*      
\end{array}
\\ \hline
\begin{array}{cc}
\vM_\vQ \vsigma & 0  \\
0 & -\vM_\vQ \vsigma^*      
\end{array}
&
\begin{array}{cc}
\xi_{\mathbf{k + Q}} & \Delta_{\mathbf{k+Q}} (i\sigma_y) \\
-\Delta^*_{\mathbf{-k-Q}}(i\sigma_y)  & -\xi_{\mathbf{-k-Q}} \\
\end{array}
\end{array} \right)
\label{Hmatrix}
\end{align}
\end{small}
where we `folded' the normal state band into the reduced Brillouin zone appropriate for the $(\pi,\pi)$-SDW unit cell. The $1/4$ in front comes from the particle-hole doubling of the bands for superconductivity and the $\vk, \vk +\vQ$ doubling for SDW. We do our analysis in the full Brillouin zone (FBZ) primarily to take advantage of the particle-hole symmetry, which simplifies the calculation of scattering rates in section \ref{sec:kinetic}. The Nambu state vector is 
\[
\Psi^{\dagger}_{\vk } = \left( 
c_{\vk \alpha_1}^{\dagger}, c_{-\vk \alpha_2}, c_{\vk+\vQ \alpha_3}^{\dagger}, c_{-\vk-\vQ \alpha_4} 
\right) 
\;,\quad \alpha_{1,2,3,4} = \uparrow,\downarrow \,.
\]
Each outlined block represents a $4\times4$ matrix constructed from spin up-down and particle-hole spaces, represented by Pauli matrices $\sigma_{x,y,z}$ and $\tau_{1,2,3}$ correspondingly. Diagonal blocks in the full matrix represent the `folded' superconducting bands, while off-diagonal  $4\times4$ blocks appear as result of SDW mixing of the electron states with momenta $\vk$ and $\vk+\vQ$ on `folded' bands. With this `folded' space we associate Pauli matrices  $\rho_{1,2,3}$. 
The Hamiltonian is (anti-)symmetric under particle-hole transformation by the construction due to superconductivity doubling ($K$ is complex conjugation) 
\begin{align}
\begin{split}
    & \cC \cH_{\vk,\vQ} \cC^{-1} = -  \cH_{-\vk,-\vQ}
    \;, \\ 
    & \cC = 1_\rho \otimes \tau_1 \otimes 1_\sigma K
\end{split}
\end{align}
Also, while the time-reversal symmetry is definitely broken in SDW state, because transformation 
$\mathcal{T} = 1_\rho \otimes 1_\tau \otimes (i\sigma_y) K$ reverses the magnetization direction $\vM \to -\vM$, 
a combination of time-reversal and a `gauge' transformation $c_{k+Q} \to -c_{k+Q}$, given by $\rho_3$, can still be a symmetry
\begin{align}
\begin{split}
    & \cTg \cH_{\vk,\vQ} \cTg^{-1} = \cH_{-\vk,-\vQ}
    \;,\\ 
    & \cTg = \rho_3 \otimes 1_\tau \otimes (i\sigma_y) K
\end{split}
\end{align}
provided $\xi_\vp = \xi_{-\vp}$, $\Delta_\vp^*=\Delta_\vp$ and $\vM_{-\vQ} = \vM_{\vQ}^*$. 
(The gauge transformation establishes an arbitrary phase $\varphi$ between states $\vk$ and $\vk+\vQ$: 
$c_{\vk+\vQ} \to e^{i\varphi} c_{\vk+\vQ}$, which results in a `slide' of the SDW profile 
$\vm(\vr) \propto Re( \langle c^\dag_{\vk+\vQ} c_{\vk} \rangle e^{-i\vQ\vr}): \quad \vM \cos(\vQ\cdot\vr) \to \vM \cos(\vQ\cdot\vr + \varphi)$, signifying arbitrariness of the coordinate origin. 
Shift by half-wavelength of SDW order, for $\varphi=\pi$, reverses the magnetisation direction, canceling the time-reversal.) 

With real $\Delta_\vk,M$ and inversion-symmetric $\xi_\vk$, both charge conjugation and `time-gauge' symmetries are present, so we split the full Hamiltonian into two independent $4\times4$ blocks for particular spin orientations $\sigma=\pm 1 (\uparrow,\downarrow)$,  
\begin{eqnarray}
\begin{aligned}
&H^{(\sigma)}=\frac{1}{4}\sum\limits_{\vk \in FBZ} \Psi^{\dagger}_{\vk \sigma} \cH^{(\sigma)}_{\vk } \Psi_{\vk \sigma}    
\\
&\cH^{(\sigma)}_{\vk } =
\begin{pmatrix}
\xi_{\vk } & \sigma\Delta_{\vk }  & \sigma M & 0 \\
\sigma\Delta_{-\vk } & - \xi_{-\vk } & 0 & \sigma M \\
\sigma M & 0 & \xi_{\vk + \vQ} & \sigma \Delta_{\vk+\vQ } \\
0 & \sigma M & \sigma \Delta_{-\vk-\vQ}  & -\xi_{-\vk-\vQ} \\
\end{pmatrix}
\label{ham matrix}
\end{aligned} 
\end{eqnarray}
where  
$\Psi^{\dagger}_{\vk \sigma}=\left( 
c_{\vk \sigma}^{\dagger}, c_{-\vk -\sigma},c_{\mathbf{k+Q}\sigma}^{\dagger},c_{\mathbf{-k-Q}-\sigma} 
\right)$ 
represents partial Nambu vector. 
Hamiltonian (\ref{ham matrix}) is diagonalized by the Bogoliubov transformation 
\begin{equation}
    \Psi_{\vk \sigma}=\begin{pmatrix} 
    c_{\vk \sigma} \\ c^\dag_{-\vk -\sigma} \\ c_{\vk +\vQ \sigma} \\ c^\dag_{-\vk-\vQ -\sigma} 
\end{pmatrix}
 = \hat B_\sigma(\vk)
 \begin{pmatrix} 
    a_{1 \vk} \\ a^\dag_{3 \vk} \\ a_{2 \vk} \\ a^\dag_{4 \vk} 
\end{pmatrix}
\label{bt}
\end{equation}
with the matrix $\hat B_\sigma(\vk)$, whose columns are the eigenvectors of the Hamiltonian matrix
\begin{equation}
    \cH^{(\sigma)}_{\vk } \hat B_\sigma(\vk) = \hat B_\sigma(\vk) \hat E_\sigma(\vk)
\end{equation}  
where
\begin{equation}
    \hat E_\sigma(\vk) = 
    \begin{pmatrix}
    E^\sigma_1(\vk) & 0 & 0 & 0 \\
    0 & -\tilde E^\sigma_1(\vk) & 0 & 0 \\
    0 & 0 & E^\sigma_2(\vk) & 0 \\
    0 & 0 & 0 & -\tilde E^\sigma_2(\vk) 
    \end{pmatrix}
\end{equation}
The $-\sigma$ spin sector is diagonalized in a similar way, and we obtain another matrix $\hat B_{-\sigma}(\vk)$, and the quasiparticle creation-annihilation operators 
\begin{eqnarray}
\Psi_{\vk -\sigma} = \hat B_{-\sigma}(\vk) \left(a_{1' \vk}, a^\dag_{3' \vk}, a_{2' \vk} , a^\dag_{4' \vk} \right)^T \,
\\
\cH^{(-\sigma)}_{\vk } \hat B_{-\sigma}(\vk) = \hat B_{-\sigma}(\vk) \hat E_{-\sigma}(\vk)
\end{eqnarray}
The two spin sectors are connected by the present symmetries. 
For example, the particle-hole transformation connects  
$\cC \cH^{(\sigma)}_{\vk } \cC^{-1} = - \cH^{(-\sigma)}_{-\vk }$ and we can identify 
$\tilde E_{1,2}^\sigma(\vk) = E_{1,2}^{-\sigma}(-\vk)$, relating eigenvectors $\hat B_{-\sigma}(-\vk) = \cC \hat B_{\sigma}(\vk)$ and the quasiparticle branches $3\vk = 1'(-\vk)$, $4\vk = 2'(-\vk)$, etc. 
The time-reversal + gauge combination reduces distinct energy levels further, by requiring $E_{1,2}^\sigma(\vk) = E_{1,2}^{-\sigma}(-\vk)$, leaving just 2 different energy values, for 4 quasiparticle branches.

Using definitions $\xi^{\pm}_{\vk }= \frac{1}{2}(\xi_{\vk } \pm \xi_{\mathbf{k+Q}})$ and $\Delta^{\pm}_{\vk }=\frac{1}{2}(\Delta_{\vk } \pm \Delta_{\mathbf{k+Q}})$, these eigenvalues of $\hat{\cH}_{\vk }$ can be written as \cite{AV, Fernandes}
\begin{align}
\begin{split}
&E^{2}_{1}(\vk)= \Gamma_{\vk } + 2\Lambda_{\vk } \quad , \quad E^{2}_{2}(\vk)= \Gamma_{\vk } - 2\Lambda_{\vk }  \\
&\Gamma_{\vk }=  (\xi^{+}_{\vk })^2 + (\xi^{-}_{\vk })^2 + (\Delta^{+}_{\vk })^2 + (\Delta^{-}_{\vk })^2 +M^2 \\
&\Lambda_{\vk }= \bigg[\bigg(\xi^{+}_{\vk }\xi^{-}_{\vk } + \Delta^{+}_{\vk }\Delta^{-}_{\vk }\bigg)^2 + M^2\bigg((\xi^{+}_{\vk })^2  +(\Delta^{+}_{\vk })^2 \bigg)\bigg]^{\frac{1}{2}}
\label{eigen}
\end{split}
\end{align}

\begin{figure}[H]
\begin{center}
\includegraphics[width=\linewidth]{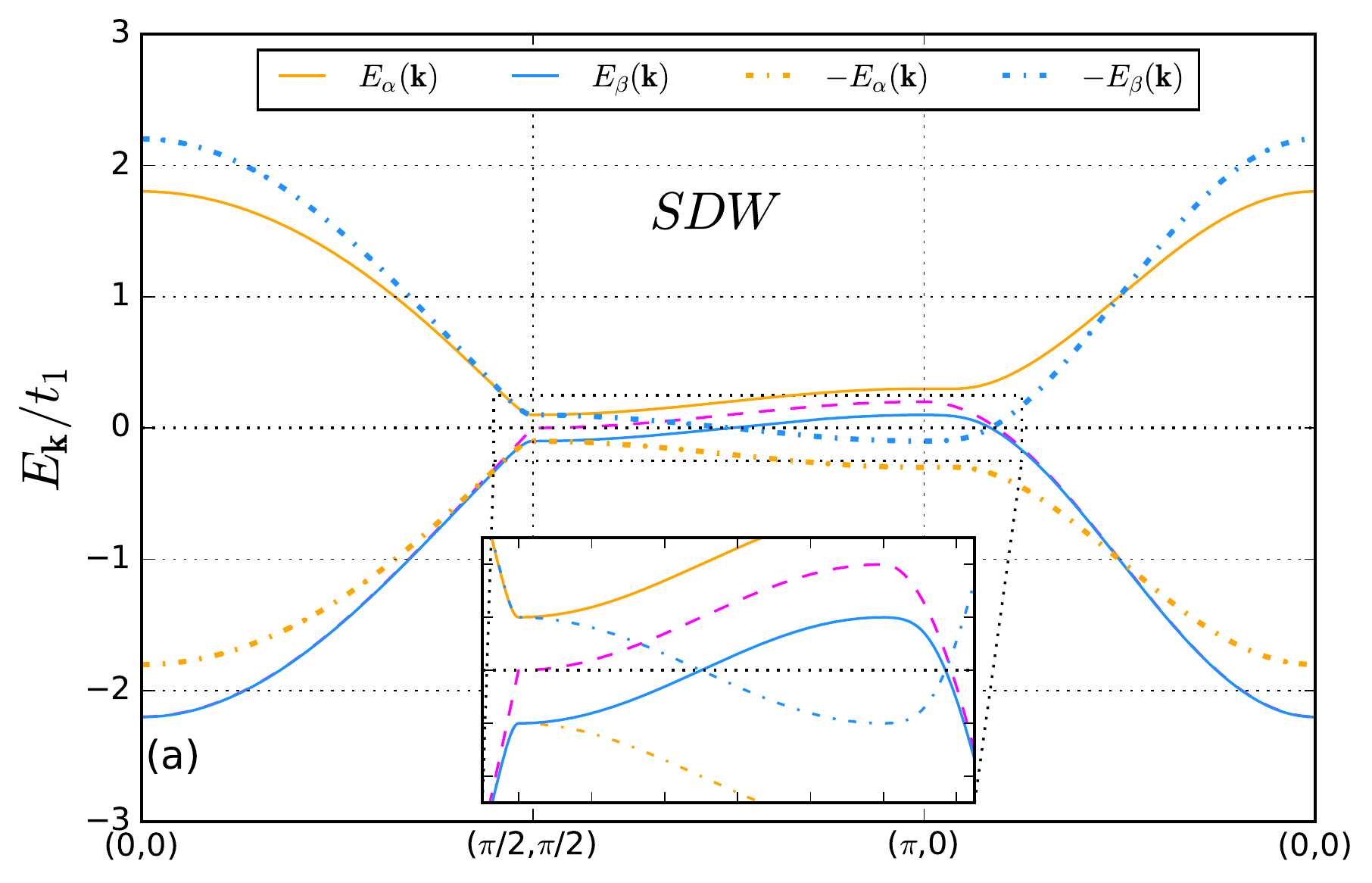}
\includegraphics[width=\linewidth]{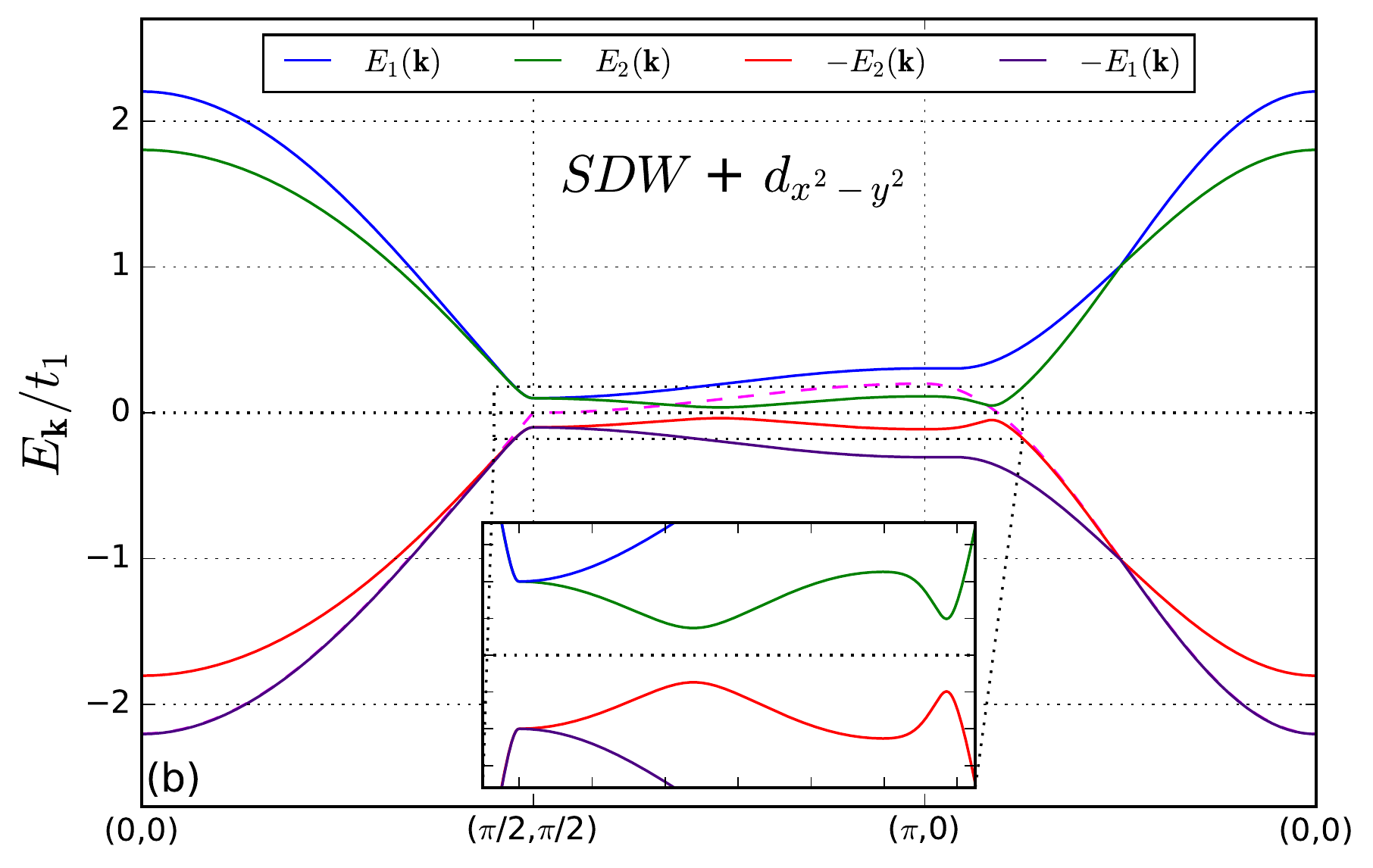}
\includegraphics[width=\linewidth]{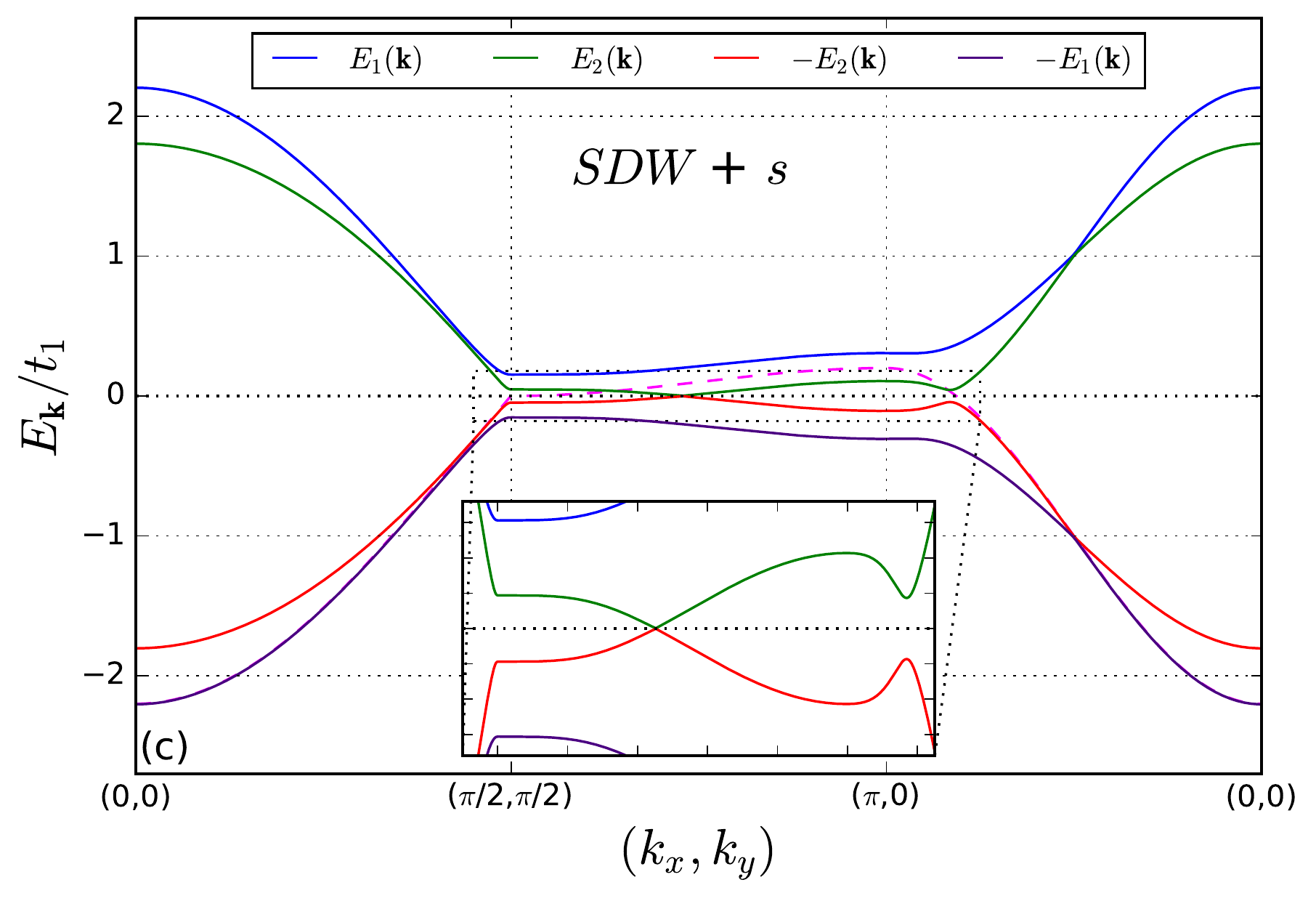}
\caption{{The quasiparticle energies in the Brillouin zone along the path $(0,0) \rightarrow(\pi/2,\pi/2) \rightarrow(\pi,0) \rightarrow(0,0)$. The normal state band is depicted by the dashed magenta curve. 
(a) the two bands $E_{\alpha, \beta}$, and their negatives, in the pure SDW phase; 
(b) the four quasiparticle bands $\pm E_{1,2}(\vk)$ in the coexisting phase $d_{x^2-y^2}$-SC and SDW; 
(c) the four quasiparticle bands in the coexisting phase $s$-SC and SDW.
Insets show the zoomed low energy sector. In the SDW state there remains a hole FS pocket around $(\pi,0)$. This remaining Fermi surface is completely gapped by emerging $d_{x^2-y^2}$-SC order. 
In the co-existing SDW and $s$-SC a Dirac nodal point remains on the boundary of RBZ. 
The parameters used for illustration are $t_2/t_1=0.2$, $M/t_1=0.1$, and $\Delta/t_1=0.05$ 
(the characteristic computed values are $M/t_1 \sim \Delta/t_1 \sim 10^{-3}$).}}
\label{bs}
\end{center}
\end{figure}
\begin{figure}[t]
\begin{center}
\includegraphics[width=\linewidth]{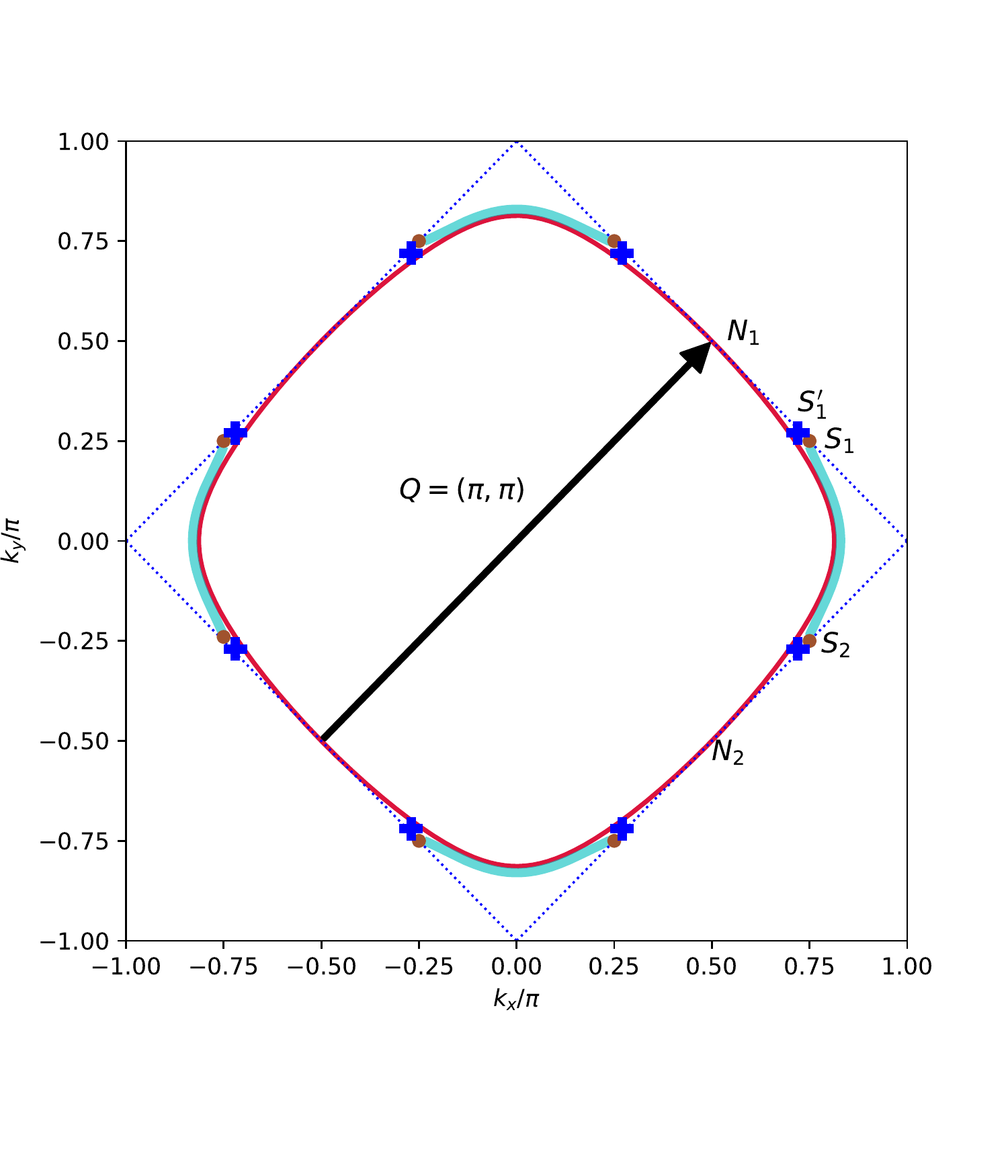}
\caption{Location of the extra nodes $S_1'$ on the boundary of the RBZ in the co-existing SDW and $s$-SC states shown by the blue crosses. They are the remnants of the FS points $S_1$ in the pure SDW state. The parameters used for illustration are $t_2/t_1=0.2$, $M/t_1=0.1$, and $\Delta/t_1=0.05$ (the characteristic computed values are $M/t_1 \sim \Delta/t_1 \sim 10^{-3}$).}
\label{FS_n}
\end{center}
\end{figure}

In the pure SDW state ($\Delta=0$) we get 
$E^{2}_{1,2}(\vk)=\left(\xi_{\vk }^{+} \pm \sqrt{(\xi_{\vk }^{-})^2 + M^2}\right)^2$ and we assign specific roots to the SDW branches as $E_{\alpha, \beta}=\xi_{\vk }^{+} \pm \sqrt{(\xi_{\vk }^{-})^2 + M^2}$ (Greek indices refer to signs $\alpha(+),\; \beta(-)$). In Fig.~\ref{bs}(a) we show the structure of the two distinct (spin degenerate) quasiparticle bands in the pure SDW phase when the FS is not perfectly nested, leaving a hole pocket $E_\beta(\vk)=0$ around $(\pi,0)$. 

In the coexistence phase we specify eigenvalues (\ref{eigen}) for the two symmetry classes: 
\textit{I=(E, O)} class ($\Delta^+_{\vk}=0$, or $\Delta_{\vk+\vQ}=-\Delta_{\vk}$), 
and \textit{II=(E, E)} class ($\Delta^-_{\vk}= 0$, or $\Delta_{\vk+\vQ}=\Delta_{\vk}$),
\begin{align}
\begin{split}
&E^{2}_{1,2;\, I,II}= \Gamma_{\vk } \pm 2\Lambda^{I,II}_{\vk }, \\ 
&\Gamma_{\vk }=   (\xi^{+}_{\vk })^2 + (\xi^{-}_{\vk })^2 + (\Delta_{\vk })^2 +M^2 \\
&\Lambda^{I}_{\vk }= \bigg[(\xi^{+}_{\vk }\xi^{-}_{\vk } )^2 + M^2(\xi^{+}_{\vk })^2\bigg]^{\frac{1}{2}} \\
&\Lambda^{II}_{\vk }=\bigg[(\xi^{+}_{\vk }\xi^{-}_{\vk })^2 + M^2\bigg((\xi^{+}_{\vk })^2  +(\Delta_{\vk })^2 \bigg)\bigg] ^{\frac{1}{2}}
\label{eigen II}
\end{split}
\end{align}
These dispersion relations have distinctly different characteristics. Spectrum of class I is completely gapped: the lowest energy state 
\begin{align}
\begin{split}
E^2_{2;I}= \bigg[\xi_{\vk }^{+} -\sqrt{(\xi_{\vk }^{-})^2 + M^2}\bigg]^2 + \Delta_{\vk}^2
\end{split}
\end{align}
can only be zero when both terms on the RHS are zero, i.e. when nodal lines of $\Delta_{\vk}$ intersect Fermi surface in the SDW state, which is impossible in this case. 
The quasiparticle bands in SDW + $d_{x^2-y^2}$-SC state are shown in Fig.~\ref{bs}(b). 

In the case of \textit{(E, E)} class (II), the spectrum has symmetry nodes on the SDW Fermi surface for $d_{xy}$ state. 
But there is also an additional nodal point on the boundary of the RBZ (where $\xi^{-}_{\vk}=0$) that is not removed by the SC order: 
\begin{align}
\begin{split}
\xi^{-}_{\vk}=0: \quad E_{2; II} = M - \sqrt{(\xi_{\vk }^{+})^2  + \Delta_{\vk}^2}=0 \,.
\label{newnode}
\end{split}
\end{align}
The nodal point, given by condition $\xi_{\vk }^{+} = \xi_\vk = \sqrt{M^2 - \Delta_{\vk}^2}$, is robust even in the SC state as long as $\Delta_\vk<M$. This point is the base of a (anisotropic) Dirac cone, obvious in the inset of Fig.~\ref{bs}(c). 
In our model location of the extra node is given by 
\begin{align}
\begin{split}
(k_x,k_y)=\frac{\pi}{2} \pm \arcsin \left[ \frac{\sqrt{M^2-\Delta^2}}{t_2} \right]^{1/2}
\end{split}
\end{align} 
and at the corresponding symmetry points, as shown by the blue crosses in Fig.~\ref{FS_n}

As a final remark, we have diagonalized the full Hamiltonian (\ref{ham matrix}) in the RBZ by computing the $\hat B_\sigma(\vk)$ eigenvectors numerically. The Hamiltonian can also be diagonalized by a `2-step process' which is sometimes employed in literature \cite{Ismer}. The methods are equivalent, we explain this in Appendix \ref{appA}. 

\subsection{Self-consistent equations for SC and SDW}
\label{sec:selfconsistency}

We solve for the mean-fields $M,\; \Delta$ self-consistently, using the Green's function method.\cite{machida2}
In the reduced Brillouin zone, the  Green's functions required to derive the self consistent equations for $\Delta_{\vk}$ are 
\[
\langle T_{\tau}c_{-\vk -\sigma}^{\dagger}(\tau)c_{\vk \sigma}^{\dagger}(0)\rangle \;, 
\langle T_{\tau}c_{\mathbf{-k-Q}-\sigma}^{\dagger}(\tau)c_{\mathbf{k+Q}\sigma}^{\dagger}(0)\rangle
\] 
and for $M$ 
\[
\langle T_{\tau} c_{\mathbf{k}\sigma}(\tau)c^{\dagger}_{\vk +\vQ \sigma}(0)\rangle \;, 
\langle T_{\tau}c^{\dagger}_{\mathbf{-k-Q}-\sigma}(\tau)c_{-\vk -\sigma}(0)\rangle \,.
\] 
They are all elements of the following bare Matsubara Green's function, which we define to be the following  $4 \times 4$ matrix, 
\begin{align}
\hat{G}(\vk ,\tau)_{ab} = -\langle T \Psi_{\vk\sigma,a}(\tau)\Psi^{\dagger}_{\vk\sigma,b}(0)\rangle
\end{align}
where the indices $a,b$ represent the components of the partial Nambu vector $\Psi^{\dagger}_{\vk \sigma, a}=\left( 
c_{\vk \sigma}^{\dagger}, c_{-\vk -\sigma},c_{\mathbf{k+Q}\sigma}^{\dagger},c_{\mathbf{-k-Q}-\sigma}\right)$. The Green's functions relevant for SC are contained in the diagonal blocks, whereas those relevant for the SDW are contained in the off-diagonal blocks. To obtain them we use the fact that $\hat{G}(\vk , \omega_n)$  satisfies the following Dyson equation
\begin{align}
\hat{G}(\vk , \omega_n)=(i\omega_n-&\hat{\cH}_{\vk })^{-1}
\label{dyson}
\end{align}
where $\omega_n= 2\pi T(n+\frac{1}{2})$ with integer $n$. Taking the paring interaction to be of the form $g(\mathbf{k,k'})=g\eta(\vk )\eta(\vk' )$, calculating the relevant Green's functions from the above Dyson equation and substituting them into (\ref{M&delta}), we arrive at the following self-consistent equations for the two symmetry classes. \cite{machida2}
%

The \textit{(E, O)} class: ${d_{x^2-y^2}}$ with $\Delta_{\vk}= \frac{1}{2}\Delta(\cos k_x - \cos k_y$)
\begin{small}
\begin{align}
\begin{split}
\frac{1}{g}=T\sum_{\omega_n}^{E_c}\sum_{\vk\in FBZ }\!\frac{\eta^2({\vk })}{D_{I}(\omega_n,\vk )} (\omega_n^2 \!+\! (\xi^{-}_{\vk })^2\!+\! (\xi^{+}_{\vk })^2\!+\! M^2\!+\! \Delta_{\vk }^2) \\
\frac{1}{U}=T\sum_{\omega_n}^{E_B}\sum_{\vk \in FBZ }\!\frac{1}{D_{I}(\omega_n,\vk )}(\omega_n^2\!+\! (\xi^{-}_{\vk })^2\!-\!(\xi^{+}_{\vk })^2\!+\! M^2\!+\!\Delta_{\vk }^2)\\
D_{I}(\omega_n,\vk )=(\omega_n^2 + (\xi^{-}_{\vk })^2 + (\xi^{+}_{\vk })^2 +\Delta_{\vk }^2 +M^2)\\-4(\xi^{+}_{\vk })^2\bigg((\xi^{-}_{\vk })^2 + M^2)\bigg)\\
=(\omega_n^2 + E_{1;I}^2)(\omega_n^2 + E_{2;I}^2)
\label{G_I}
\end{split}
\end{align}
\end{small}
The \textit{(E, E)} class: the isotropic s-wave with $\Delta_{\vk}= \Delta$  and  $d_{xy}$ with $\Delta_{\vk}= \Delta\sin k_x \sin k_y$
\begin{small}
\begin{align}
\begin{split}
\frac{1}{g}\!=T\!\sum_{\omega_n}^{E_c}\!\sum_{\vk \in FBZ }\!\frac{\eta^2({\vk })}{D_{II}(\omega_n,\vk )} (\omega_n^2\! +\!(\xi^{-}_{\vk })^2 \!+\!(\xi^{+}_{\vk })^2 \!-\! M^2 \!+\! \Delta_{\vk }^2) \\
\frac{1}{U}\!=T\!\sum_{\omega_n}^{E_B}\!\sum_{\vk \in FBZ }\!\frac{1}{D_{II}(\omega_n,\vk )}(\omega_n^2\! +\! (\xi^{-}_{\vk })^2 \!-\!(\xi^{+}_{\vk })^2\! +\! M^2 \!-\! \Delta_{\vk }^2)\\
D_{II}(\omega_n,\vk )=(\omega_n^2 + (\xi^{-}_{\vk })^2 + (\xi^{+}_{\vk })^2 +\Delta_{\vk }^2 +M^2)\\
-4(\xi^{+}_{\vk })^2\bigg((\xi^{-}_{\vk })^2 + M^2)\bigg) -4\Delta_{\vk }^2M^2\\
=(\omega_n^2 + E_{1;II}^2)(\omega_n^2 + E_{2;II}^2)
\label{G_II}
\end{split}
\end{align}
\end{small}
where $E_{1,2; I,II}$ are the quasiparticle energies for symmetry classes I and II, defined in (\ref{eigen II}). 
$E_C$ and $E_B$ are the SC cutoff and the SDW cutoff energies respectively.

\subsubsection{Numerical Solution of Self Consistent equations}

In the following, we solve the self-consistent equations (\ref{G_I}) and (\ref{G_II}) for band parameters $t_1/2\pi T_{SDW}=100$, $t_2/2\pi T_{SDW}=10$ , $E_C/2\pi T_{SDW} =30$ and $E_B/2\pi T_{SDW}=60$, and eliminate interactions $g$ and $U$, to obtain the temperature dependence of the order parameters $\Delta(T)$ and $M(T)$. 

For perfect nesting with $t_2=0$, SDW order gaps the entire FS. This prohibits SC order to open up a gap anywhere on the FS when $T_{C0} < T_{SDW}$, and the superconductivity never appears in this case. 
This is verified by numerically solving the self-consistent equations (\ref{G_I}) and (\ref{G_II}) with $t_2=0 $. The result is the usual BCS-profile for $M(T)$ and $\Delta=0$. 

However, both SC and SDW orders can appear when we go away from the perfect nesting limit. The nature of the coexistence is very different for the \textit{(E, O)} and \textit{(E, E)} symmetry classes. 
Solutions of the self-consistent equations (\ref{G_I}) and (\ref{G_II}) depend of the parameter $p=\frac{T_{C0}}{T_{SDW}}$. 
We show the order parameter profiles later, together with thermal conductivity results, in section \ref{sec:results} and here summarize the main points.  

For the \textit{(E, O)} class of SDW+${d_{x^2-y^2}}$-SC, one numerically solves Eqs.~(\ref{G_I}). In this case SC can naturally co-exist with SDW, and in fact below $T_C$ the magnetization $M$ is enhanced compared to the pure SDW state. SC transition temperature is also increased on the SDW background, $T_C > T_{C0}$, 
see Figs.~\ref{k_d22}(a,c). 
The SDW\textrightarrow SDW+SC transition is always second-order.

For states of \textit{(E, E)} class the interplay is more complicated. In Fig.~\ref{k_dxy}(a,c) we show numerical solution of (\ref{G_II}) for the $d_{xy}$ pairing state. 
Depending on the value of $T_{C0}/T_{SDW}$ the SDW\textrightarrow SC transition can be either first- or second-order. 
For relatively strong SC order, $p=0.5$, the SC state completely replaces SDW order via a first-order transition. For a lower $p=0.35$, SDW survives and allows for a smaller $\Delta$-order to appear simultaneously through a second-order transition.
This competition comes with suppression of the superconducting transition temperature in the presence of the SDW background,  $T_C<T_{C0}$. 
Behavior for the isotropic $s$-wave state is similar to $d_{xy}$ case, Fig.~\ref{k_s}.

\subsection{Kinetic Method for Heat Conductivity}
\label{sec:kinetic}

We use the Boltzmann kinetic-equation approach to calculate the thermal conductivity for the system with intertwined orders. This method was widely used to compute to compute thermal conductivity, both in s-wave superconductor\cite{bardeen,gel}, as well as in unconventional superconductors\cite{mineev,arfi,arfi2,fritz}, and for quantum critical systems \cite{damle,sachdev,senthil}. We begin with the expression of the total heat current carried by the quasiparticles

\begin{align}
\mathbf{j}_E=2\sum_{n=1}^2 \mathbf{j}_{n}=2\sum_{n=1}^2\intk E_n(\vk ) \mathbf v_n(\vk )f_n(\vk )
\label{heat_flux}
\end{align} 
{In the above expression we integrate momentum over the FBZ, double-counting the states, and therefore requires an extra factor of 2 in denominator: $2 \times 4\pi^2$.
}
The sum is over the two quasiparticle branches with distinct energies  $E_{1}$ and $E_{2}$ (as given in (\ref{eigen})) and $f_n(\vk )$ is the distribution function of the respective quasiparticle branches. The factor of two takes care of the spin degeneracy of each branch. The thermal conductivity tensor  is the proportionality coefficient between the heat current and temperature gradient.
\begin{align}
(\mathbf{j}_E)_i=-\kappa_{ij} \mathbf{{\nabla}}_j T
\label{tc}
\end{align} 
The quasi-particle distribution function  $f_n(\vk )$ satisfies the Boltzmann equation  
\begin{align}
\frac{\partial f_n(\vk)}{\partial t} + \frac{\partial E_n}{\partial \vk }\grad f_n - \grad E_n  \frac{\partial f_n(\vk)}{\partial \vk  } = I^{coll}_n(\vk )
\label{boleqn} 
\end{align} 
$I(\vk )$ being the collision integral. We follow the usual process of linearizing the left hand side of (\ref{boleqn}) by writing $f_n(\vk) =f_n^{0}(\vk) + \delta f_n(\vk)$, where $f^{0}_n(\vk)= \frac{1}{e^{E_{n}( \vk)/T}+1}$ is the equilibrium Fermi-Dirac distribution function. $\delta f_n(\vk)$ is the deviation from the equilibrium value caused by the presence of the stationary thermal gradient. The linearization yields \cite{mineev},
\begin{align}
\frac{\partial \delta f_n(\vk )}{\partial t} -E_{n}(\vk ) \mathbf{v}_n(\vk )\frac{\grad T}{T}\frac{\partial f^0_n(\vk ) }{\partial E_n}=I^{coll}_n(\vk )
\label{lboleqn} 
\end{align} 

The quasiparticle velocity is defined as 
\begin{align}
\mathbf{v}_n(\vk)=\nabla_{\vk}E_n(\vk)
\label{qpvel} 
\end{align} 
For a stationary thermal gradient the first term is zero. We now look at the right hand side, the collision integrals in the case of weak disorder is obtained my multiplying the contribution of a single impurity by their concentration $N_{imp}$: 
\begin{widetext}
\begin{align}
\begin{split}
I^{coll}_1(\vk )&= N_{imp}\intkprfbz\bigg[W_{11}(\vk ,\vk' ) \bigg(\delta f_1(\vk' ) - \delta f_1(\vk )\bigg)
+W_{12}(\vk ,\vk' ) \bigg(\delta f_2(\vk' ) - \delta f_1(\vk )\bigg)\bigg]\\
I^{coll}_2(\vk )&= N_{imp}\intkprfbz\bigg[W_{22}(\vk ,\vk' ) \bigg(\delta f_2(\vk' ) - \delta f_2(\vk )\bigg)
+W_{21}(\vk ,\vk' ) \bigg(\delta f_1(\vk' ) - \delta f_2(\vk )\bigg)\bigg]\\
\label{CI} 
\end{split}
\end{align} 
where $W_{nm}(\vk ,\vk' )$ is the rate of elastic scattering between the quasiparticle branches $n$ and $m$ in the FBZ. Therefore we have two coupled kinetic equations for $\delta f_1(\vk )$ and $\delta f_2(\vk )$. We can rewrite (\ref{lboleqn}) for $\delta f_n(\vk )$ as 

\begin{align}
\begin{split}
E_{1}(\vk ) \mathbf{v}_{1}(\vk )\frac{\nabla T}{T} \frac{\partial f^0_{1}(\vk )}{\partial E_{1}}=-N_{imp}\bigg[ \intkprfbz\bigg(W_{11}(\mathbf{k,k'})\delta f_1(\vk' ) + W_{12}(\mathbf{k,k'})\delta f_2(\vk' )\bigg)\bigg] + \bigg(\frac{1}{\tau_{11}} + \frac{1}{\tau_{12}}\bigg)\delta f_{1}(\vk )\\
E_{2}(\vk ) \mathbf{v}_{2}(\vk )\frac{\nabla T}{T} \frac{\partial f^0_{2}(\vk)}{\partial E_{2}}=-N_{imp}\bigg[ \intkprfbz\bigg(W_{22}(\mathbf{k,k'})\delta f_1(\vk' ) + W_{21}(\mathbf{k,k'})\delta f_2(\vk' )\bigg)\bigg] + \bigg(\frac{1}{\tau_{22}} + \frac{1}{\tau_{21}}\bigg)\delta f_{2}(\vk )
\end{split}
\label{ckeqn}
\end{align}
\end{widetext} 
where we have defined the quasiparticle relaxation time as 

\begin{align}
\tau^{-1}_{nm}(\vk )=N_{imp}\intkprfbz W_{nm}(\mathbf{k,k'})
\label{scat rate}
\end{align} 
The above equations are decoupled by the usual symmetry argument\cite{mineev,arfi}. The driving term is odd under spatial inversion since $\mathbf{v(-\vk )}=-\mathbf{v(\vk )}$, whereas the quasiparticle relaxation time is even under spatial inversion $\tau^{-1}_{nm}(-\vk )= \tau^{-1}_{nm}(\vk)$ due to symmetry $W_{nm}(-\vk , -\vk')= W_{nm}(\vk,\vk')$, which implies that  $\delta f_{n}(\vk )$ is  odd under $\vk \rightarrow -\vk $. Thus the first terms on the right in (\ref{ckeqn}) are integrals of odd functions over a symmetric region of integration and therefore go to zero: 
$\big[ \dots \big]=0$, which represents vanishing vertex corrections. Thus the heat current carried by the  quasiparticle branches with energies $E_{1,2}(\vk)$ are
\begingroup
\thickmuskip=-0mu
\begin{small}
\begin{align}
\begin{split}
(\mathbf{j}_{1})_i =\intk E^2_{1}(\vk ){v}_{1i}(\vk ){v}_{1j}(\vk )\frac{\nabla_j T}{T} \frac{\partial f^0_{1}(\vk )}{\partial E_{1}}\bigg(\frac{1}{\tau_{11}} + \frac{1}{\tau_{12}}\bigg)^{-1}\\
(\mathbf{j}_{2})_i=\intk E^2_{2}(\vk ) {v}_{2i}(\vk ) {v}_{2j}(\vk )\frac{\nabla_j T}{T} \frac{\partial f^0_{2}(\vk )}{\partial E_{2}}\bigg(\frac{1}{\tau_{22}} + \frac{1}{\tau_{21}}\bigg)^{-1}
\end{split}
\label{heatcurrent}
\end{align}
\end{small}
\endgroup
which 
results in the following expression for the thermal conductivity tensor,
\begin{small}
\begin{align}
\begin{split}
\kappa_{ij}&=(\kappa_{1})_{ij} + (\kappa_{2})_{ij}\\
(\kappa_{1})_{ij}&=-\frac{2}{T}\intk E^2_{1}(\vk ) {v}_{1i}(\vk ){v}_{1j}(\vk )\frac{\partial f^0_{1}(\vk )}{\partial E_{1}}\bigg(\frac{1}{\tau_{11}} + \frac{1}{\tau_{12}}\bigg)^{-1}\\
(\kappa_{2})_{ij}&=-\frac{2}{T}\intk E^2_{2}(\vk ){v}_{2i}(\vk ){v}_{2j}(\vk )\frac{\partial f^0_{2}(\vk )}{\partial E_{2}}\bigg(\frac{1}{\tau_{22}} + \frac{1}{\tau_{21}}\bigg)^{-1}
\end{split}
\label{tc2}
\end{align}
\end{small}
The expression for the scattering rate in the Born limit is given by 
\begin{small}
\begin{align}
\begin{split}
W_{nm}(\mathbf{k,k'})&=\frac{2\pi}{\hbar} 
|\langle\vk' ,n|H_{imp}|\vk ,m\rangle |^2\delta (E_n(\vk )-E_m(\vk' ))\\
&=\frac{2\pi}{\hbar}|V(\mathbf{k,k'})|^2 |C_{nm}(\mathbf{k,k'})|^2\delta (E_n(\vk )-E_m(\vk' ))
\label{golden}
\end{split} 
\end{align}
\end{small} 
The quasiparticle state with momentum $\vk$ and energy $E_{1}(\vk)$ is defined as $|\vk ,1\rangle= a^{\dagger}_{1\vk}|0\rangle $. Similarly quasiparticle state with momentum $\vk$ and energy $E_{2}(\vk)$ is defined as $|\vk ,2\rangle= a^{\dagger}_{2\vk}|0\rangle$. $|0\rangle$ is the vacuum state with no quasiparticles. $\langle\vk' ,n|H_{imp}|\vk ,m\rangle$ is the amplitude for a single impurity to scatter from the particle state $\vk $ with energy $E_m(\vk )$ to the state $\vk' $ with energy $E_n(\vk' )$. The matrix $C_{nm}(\mathbf{k,k'})$ contains the coherence factors coming from the Bogoliubov transformation between the normal and ordered states. In the following we consider the case of an isotropic scattering amplitude $V(\mathbf{k,k'}) = V =$ const. Therefore the expression (\ref{scat rate}) for the quasiparticle lifetimes become,
\begin{small}
\begin{align}
\tau^{-1}_{nm}(\vk )=N_{imp}V^2\frac{2\pi}{\hbar}\intkprfbz |C_{nm}(\mathbf{k,k'})|^2\delta (E_i(\vk )-E_j(\vk' ))
\label{scat rate2}
\end{align} 
\end{small}  
\begin{figure}[t]
\centering
\includegraphics[width=9cm, height=6cm,angle=0]{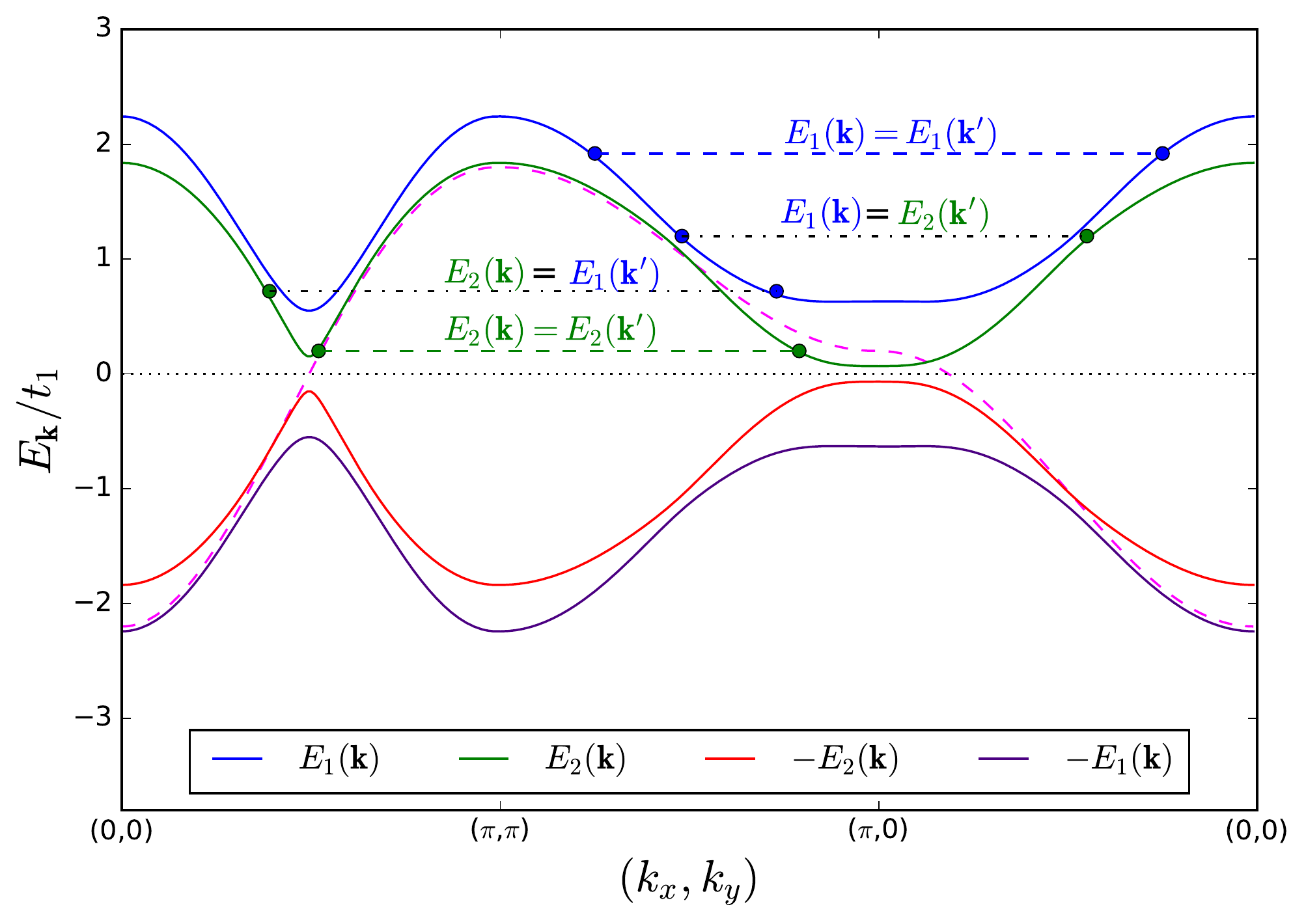}
\caption{\label{bs_scat} Schematic representation of the various scattering processes in the full BZ.  The Band structure is  along the path $(0,0) \rightarrow(\pi,\pi) \rightarrow(\pi,0) \rightarrow(0,0)$. The dashed blue and green horizontal lines represent intra-band scattering processes, the dash-dotted black horizontal lines represent inter-band scattering processes. The parameters used for illustration are $t_2/t_1=0.2$, $M/t_1=0.35$. and $\Delta/t_1=0.2$.
}
\end{figure}

We evaluate the momentum integral (\ref{scat rate2}) numerically using the high precision sampling method \cite{pax}. Using $\tau^{-1}_{nm}(\vk )$ from (\ref{scat rate2}) we numerically evaluate the momentum integrals in (\ref{tc2}) over the FBZ, see Fig.~\ref{bs_scat}. We also numerically compute the  values for  $\tau^{-1}_{nm}(\vk )$ and $\kappa(T)$ in the normal state, by setting $\Delta=0$ and $M=0$ in equations (\ref{scat rate2}) and (\ref{tc2}) respectively, and eliminate the unknown $N_{imp}V^2$ in favor of normal state relaxation time $\tau_N$ that only appears in $\kappa_N(T)$. We assume $\tau_N^{-1}$ is small enough and neglect order  parameter suppression by impurities.  
%
%
The matrix of coherence factors $C_{nm}(\mathbf{k,k'})$ is also computed numerically, by first writing the impurity scattering Hamiltonian in the same Nambu basis as (\ref{ham matrix}) 

\begin{align}
\begin{split}
H_{imp}&=V\sum_{\mathbf{k,k'},\sigma} c_{\vk' \sigma}^{\dagger}c_{\vk \sigma}\\
&=\frac{V}{4}\sum\limits_{\mathbf{k,k'} \in FBZ} \Psi^{\dagger}_{\vk'\sigma,a}\mathcal{S}_{ab}\Psi_{\vk\sigma,b}    \\
\mathcal{S}_{ab} &=
\begin{pmatrix}
1& 0  & 0 & 0 \\
0 & - 1 & 0 & 0 \\
0 & 0 & 1 & 0 \\
0 & 0 & 0  & -1 \\
\end{pmatrix}
\label{himp}
\end{split} 
\end{align}
where the factor $\frac{1}{4}$ comes from both particle-hole doubling for SC and $(\vk, \vk+\vQ) $ doubling  for SDW in the FBZ. 
Upon performing the Bogoliubov transformation (\ref{bt}) on the Nambu vectors we get, 
\begin{equation}
H_{imp}=\frac{V}{4}\sum\limits_{\mathbf{k,k'} \in FBZ} A^{\dagger}_{\vk'a}D_{ab}(\vk,\vk')A_{\vk,b} 
\end{equation}
where $A^{\dagger}_{\vk}=\left(a_{1\vk}^{\dagger}, a_{3\vk},a_{2\mathbf{k}}^{\dagger},a_{4\mathbf{k}}\right)$ and the matrix $\hat{D}(\vk,\vk')$ from which we get the coherence factors
\begin{align}
\hat{D}(\vk,\vk') = \hat{B}^{\dagger}_{\sigma}(\vk') \mathcal{\hat{S}} \hat{B}_{\sigma}(\vk) 
\label{cmatrix}
\end{align}    
The $\vk $ dependence in  $\hat{D}(\mathbf{k,k'})$ comes from $\Delta(\vk )$ and $\xi(\vk )$ through  the eigenvectors of $\mathcal{\hat{H}}_{\vk }$ and since we artificially quadruple our bands we only include physically available in-band scattering, so $\mathcal{S}_{ab}$ is diagonal. From the ordering of the $A^{\dagger}_{\vk}$-vector, 
the intra-band coherence factors 
$$ C_{11}(\mathbf{k,k'})=D_{11}(\mathbf{k,k'}) \;, \quad C_{22}(\mathbf{k,k'})=D_{33}(\mathbf{k,k'}),$$ 
and inter-band 
$$ C_{12}(\mathbf{k,k'})=D_{13}(\mathbf{k,k'}) \;, \quad C_{21}(\mathbf{k,k'})=D_{31}(\mathbf{k,k'})$$ 
- all for scattering inside the FBZ, as shown in Fig.~\ref{bs_scat}. 

\section{Numerical Results and Discussion}
\label{sec:results}

\begin{figure}
\begin{center}
\includegraphics[width=\linewidth]{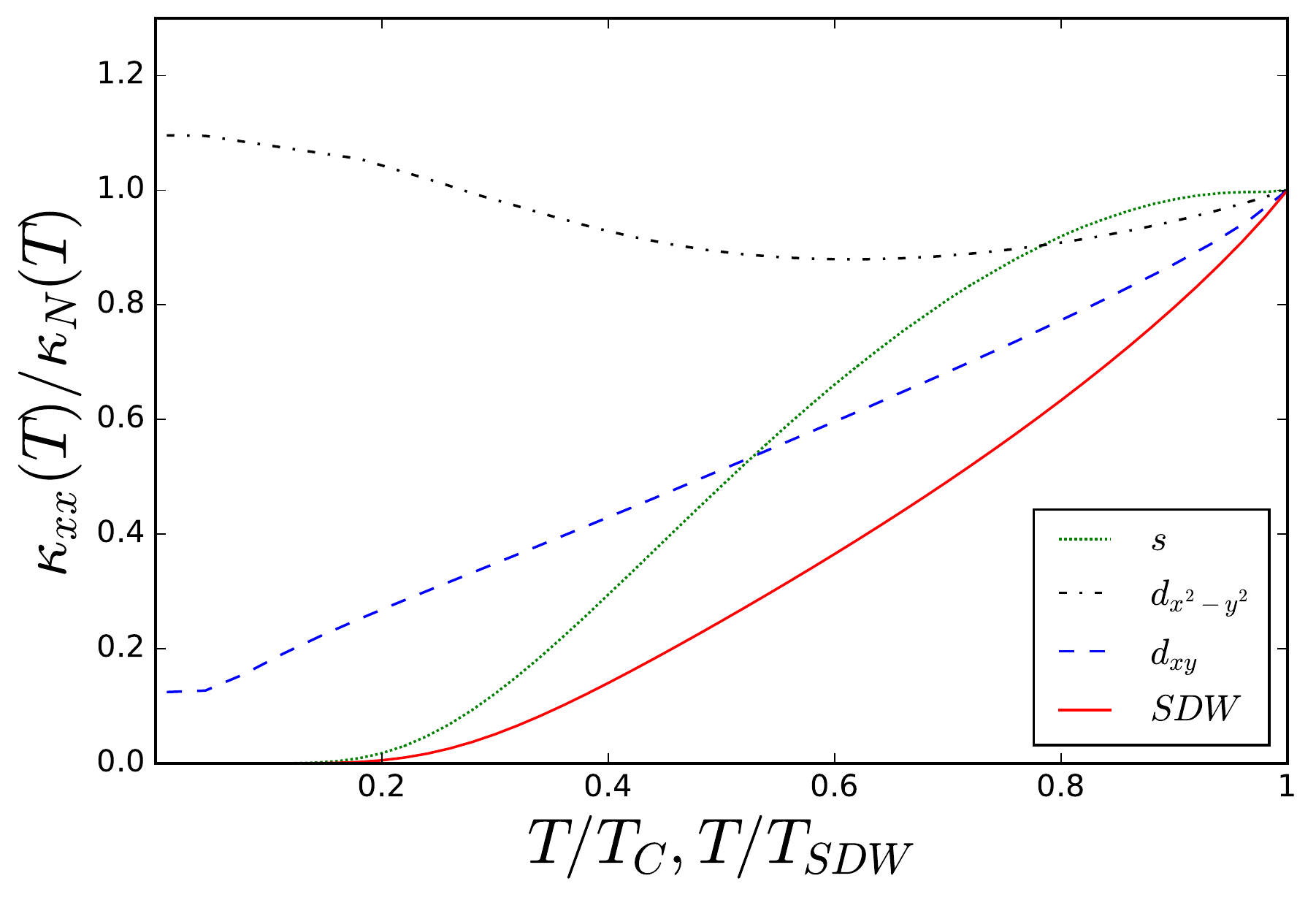}
\caption{\label{k_sc} 
Thermal conductivity for pure SC states: $s$-, $d_{x^2-y^2}$- and $d_{xy}$-wave 
(dispersion parameters $t_1/2\pi T_{C}=100$, $t_2/2\pi T_{C}=10$), 
or for pure SDW state in perfectly nested regime ($t_1/2\pi T_{SDW}=100$, $t_2/2\pi T_{SDW}=0$).
}
\end{center}
\end{figure}

We begin our discussion by first calculating thermal conductivity of the pure SC or SDW states for our tight binding model. For various pairing, $s$-, $d_{x^2-y^2}$- and $d_{xy}$-wave, the values of $\Delta(T)$ are obtained by self consistently solving the weak coupling gap equation, {and neglecting $T_c$ suppression by impurities}. The numerical results for $\kappa_{xx}$ are shown in Fig.~\ref{k_sc}. 
We see the characteristic exponential fall in the thermal conductivity for the isotropic fully-gapped $s$-wave superconductor \cite{bardeen, gel}. 
The general behavior of $\kappa(T)/T$ for the $d_{xy}$ and $d_{x^2-y^2}$ states also agrees with earlier calculations\cite{graf,arfi}, where the low-$T$ regime is dominated by the nodal quasiparticles, producing the finite residual $\kappa/T$.
However, while for circular FS $\kappa_{xx}$ is the same for the $d_{xy}$ and $d_{x^2-y^2}$ states, in the case of our anisotropic FS the two symmetries result in very different values of heat conductivity. The $d_{x^2-y^2}$ pairing has nodes on flat parts of the FS with large Fermi velocity and smaller DOS. By gapping the corners of the FS with large DOS, the scattering rate is significantly reduced, producing longer-lived high-velocity nodal quasiparticles that result in heat conductivity exceeding that of the normal state. The $d_{xy}$ state, on the other hand, has nodes where Fermi velocity is small, resulting in much lower $\kappa/T$. 

For completeness, we note that for strong scattering centers one has to go beyond Born approximation to explain experimental data in cuprates and heavy-fermions \cite{pethick,arfi}. 
{Also, quasiparticle Boltzmann approach fails at low temperatures when low-energy quasiparticles cannot be well-established due to impurity broadening\cite{arfi}.}

We also show thermal conductivity for pure SDW state 
when the Fermi surface is nested perfectly i.e. $\mu=0$ and $t_2=0$ in our model.  
The SDW opens a gap along entire FS. The sharp fall in the thermal conductivity seen in Fig.~\ref{k_sc} is often seen in thermal conductivity experiments on spin density wave antiferromagnets\cite{kim2,sayles,steckel}.

The difference between slopes of $\kappa(T)$ just below the transition temperature for $s$-, $d$-wave superconductors and SDW can be explained by the difference in coherence factors. 
For a singlet superconductor,
\begin{eqnarray}
&& |C^{SC}_{11}|^2\!=\! \frac{1}{2}\left(1+\frac{\xi_{\vk }\xi_{\vk' }-\Delta_{\vk }\Delta_{\vk' }}{E_1(\vk)E_2(\vk')}\right), 
\\
&& |C^{SC}_{22}|^2 = \frac{1}{2}\left(1+\frac{\xi_{\vk+\vQ }\xi_{\vk'+\vQ }-\Delta_{\vk+\vQ }\Delta_{\vk'+\vQ }}{E_2(\vk)E_2(\vk')}\right)
\;\\ 
&& |C^{SC}_{12}|^2 = |C^{SC}_{21}|^2=0
\end{eqnarray}
with 
$E_1(\vk)=\sqrt{(\xi_{\vk})^2 +\Delta^2_{\vk  }}$ and 
$E_2(\vk )=\sqrt{(\xi_{\vk + \vQ })^2 +\Delta^2_{\vk + \vQ }}$ being the two branches that together count the states of the  superconductor in the full BZ. 
For the SDW state
\begin{eqnarray}
&& |C_{1 1}^{SDW}|^2= |C_{22}^{SDW}|^2= \frac{1}{2}\left(1+\frac{\xi^{-}_{\vk }\xi^{-}_{\vk' } + M^2}{\Pi_{\vk }\Pi_{\vk' }}\right), 
\\
&& |C_{12}^{SDW}|^2= |C_{21}^{SDW}|^2=\frac{1}{2}\left(1-\frac{\xi^{-}_{\vk }\xi^{-}_{\vk' } + M^2}{\Pi_{\vk }\Pi_{\vk' }}\right)
\end{eqnarray}
$\Pi_{\vk}= \sqrt{(\xi^{-}_{\vk })^2 +M^2}$. 
{
For both the $s$ and $d$ wave superconductors, the $\xi_{\vk} \xi_{\vk'}$ terms vanish after ${\vk'}$ integration in  equation (\ref{scat rate2}), due to cancellation of positive and negative $\xi_{\vk'}$ contributions. 
This leaves 
$\tau^{-1}_{11}(\vk) = \tau_{N}^{-1}\frac{N(E_1(\vk))}{N_{0}} \left( 1-\frac{\Delta^2}{E^2_1(\vk)} \right)$  
and 
$\tau^{-1}_{22}(\vk) = \tau_{N}^{-1}\frac{N(E_2(\vk))}{N_{0}} \left(1-\frac{\Delta^2}{E^2_2(\vk)} \right)$ 
for the $s$-wave case. For the $d$-wave states also the $\Delta_{\vk}\Delta_{\vk'}$ terms vanish on integrating over the directions of $\vk'$ in equation (\ref{scat rate2}), resulting in 
$\tau^{-1}_{11}(\vk) = \tau_{N}^{-1}\frac{N(E_1(\vk))}{N_{0}}$ 
and 
$\tau^{-1}_{22}(\vk) = \tau_{N}^{-1}\frac{N(E_2(\vk))}{N_{0}}$ for the $d$-wave states. 
$N(E_{1,2}(\vk))$ denotes the density of SC states with energies $ E_{1,2}(\vk)$, 
and $N_{0}$ being the normal density of states at the Fermi level. 
For a SDW with a perfectly nested FS, 
$\xi_{\vk + \vQ}=-\xi_{\vk }$, so $\xi^{+}_{\vk }=0$ and $\xi^{-}_{\vk }=\xi_{\vk }$. 
Thus $E_{\alpha}(\vk)=\Pi_{\vk}$ and   $E_{\beta}(\vk)= -\Pi_{\vk}$. 
Again, terms $\xi^-_{\vk} \xi^-_{\vk'}$ will drop out under the $\vk'$ integration in  equation (\ref{scat rate2}) , leaving 
$\tau^{-1}_{11}(\vk) = \tau_{N}^{-1}\frac{N(E_{\alpha}(\vk))}{N_{0}}\left(1+\frac{M^2}{E^2_{\alpha}(\vk)} \right)$ and $\tau^{-1}_{22}(\vk) = \tau_{N}^{-1}\frac{N(E_{\beta}(\vk))}{N_{0}}\left(1+\frac{M^2}{E^2_{\beta}(\vk)}\right)$. 
For perfectly nested FS the two bands do not overlap in energy and thus there is no inter-band scattering,  $\tau^{-1}_{12}(\vk) = \tau^{-1}_{21}(\vk)=0$. $N(E_{\alpha,\beta}(\vk))$ once again denotes the density of SDW quasiparticle states with energy $ E_{\alpha,\beta}(\vk)$}. 


Comparing the coherence factors for various states, one can notice that the effective relaxation times in equation (\ref{tc2}) have this hierarchy near their transition temperatures 
\[
\tau_{SDW} < \tau_{d} < \tau_{s} 
\]
resulting in the observed different slopes in Fig.~\ref{k_sc}. 
Finally, we note that the difference in signs inside coherence factors for fully gapped $s$-SC ($1-\Delta^2/E^2$) and SDW ($1+M^2/E^2$), comes from the particle-hole difference in the impurity scattering matrix, 
$\mathcal{S}^{SC}_{ab} \propto diag(1,-1)$ vs $\mathcal{S}_{ab}^{SDW} \propto diag(1,1)$.

\subsection{The $(E,O)$ class}

\begin{figure}
\centering
\includegraphics[width=8cm, height=5.2cm,angle=0]{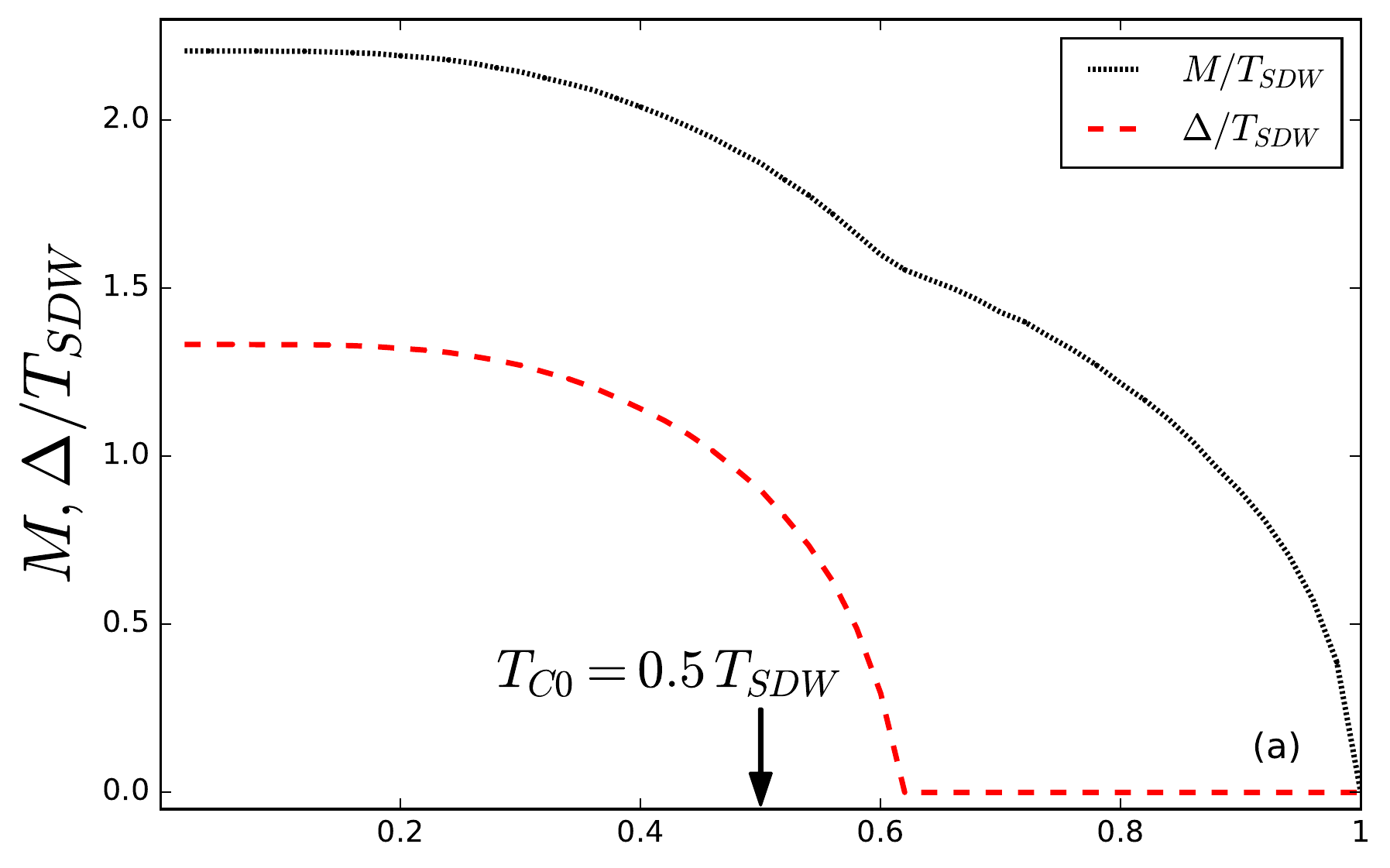}
\begin{overpic}[scale=0.47]{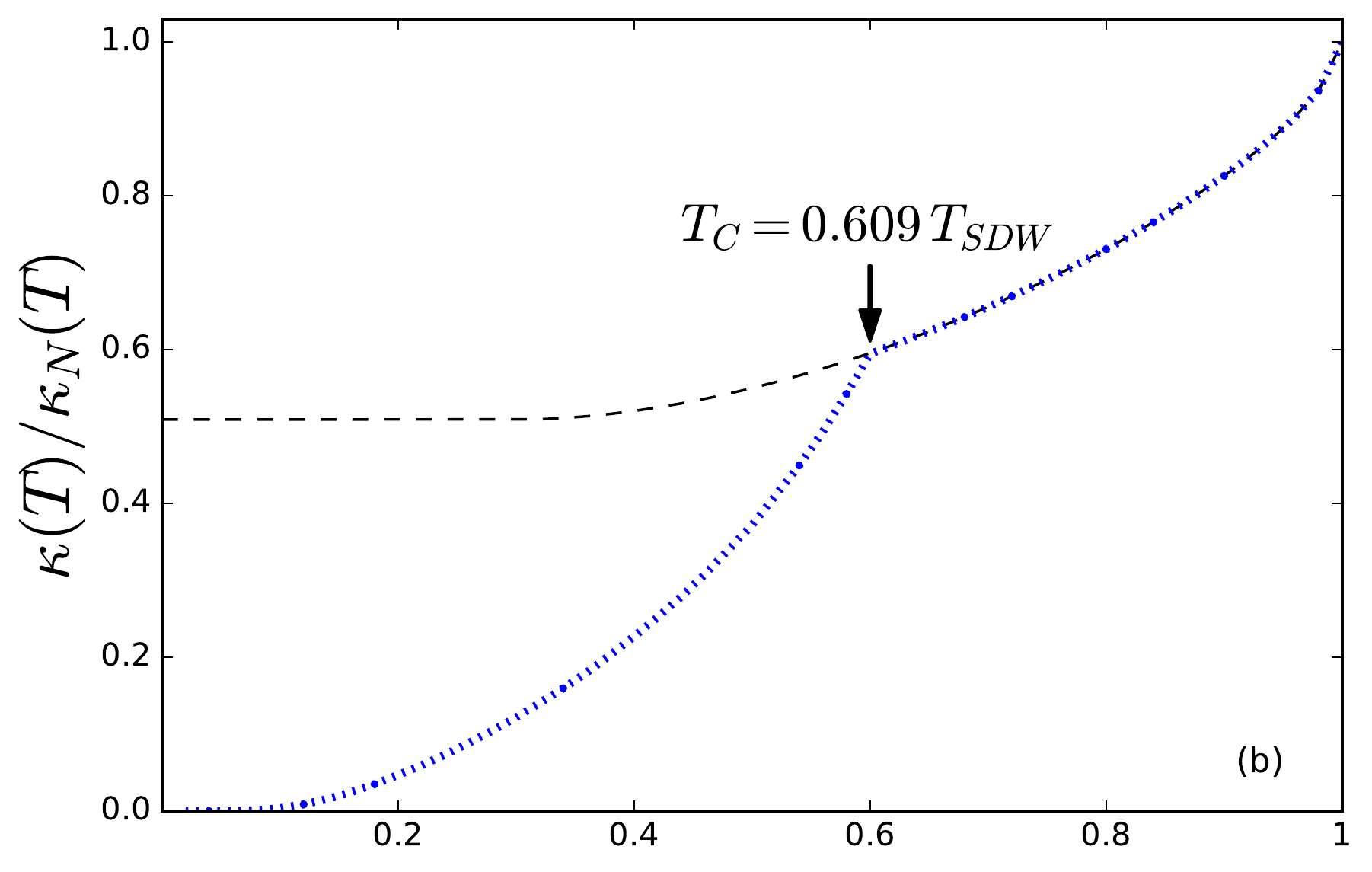}
\put(59,5){\includegraphics[scale=0.30]{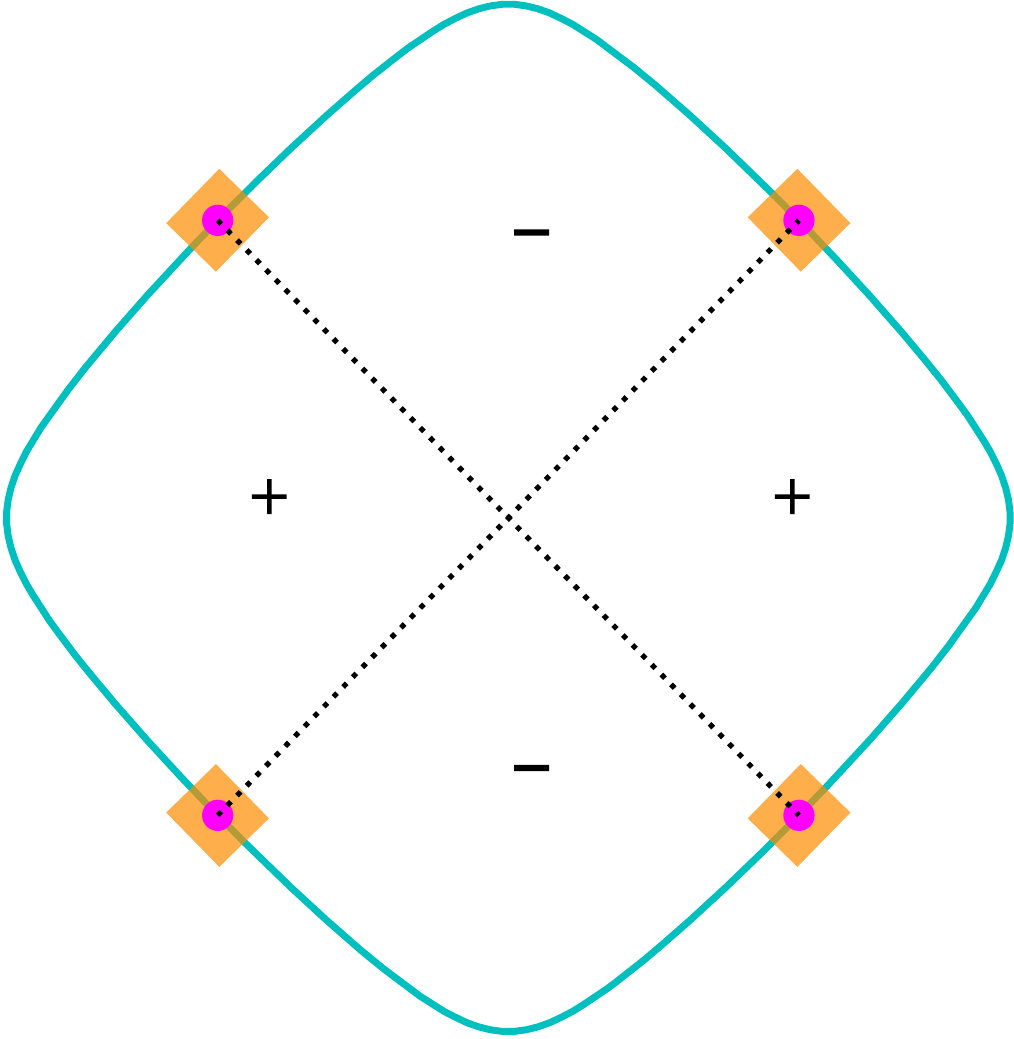}}
\end{overpic}
\includegraphics[width=8cm, height=5.2cm,angle=0]{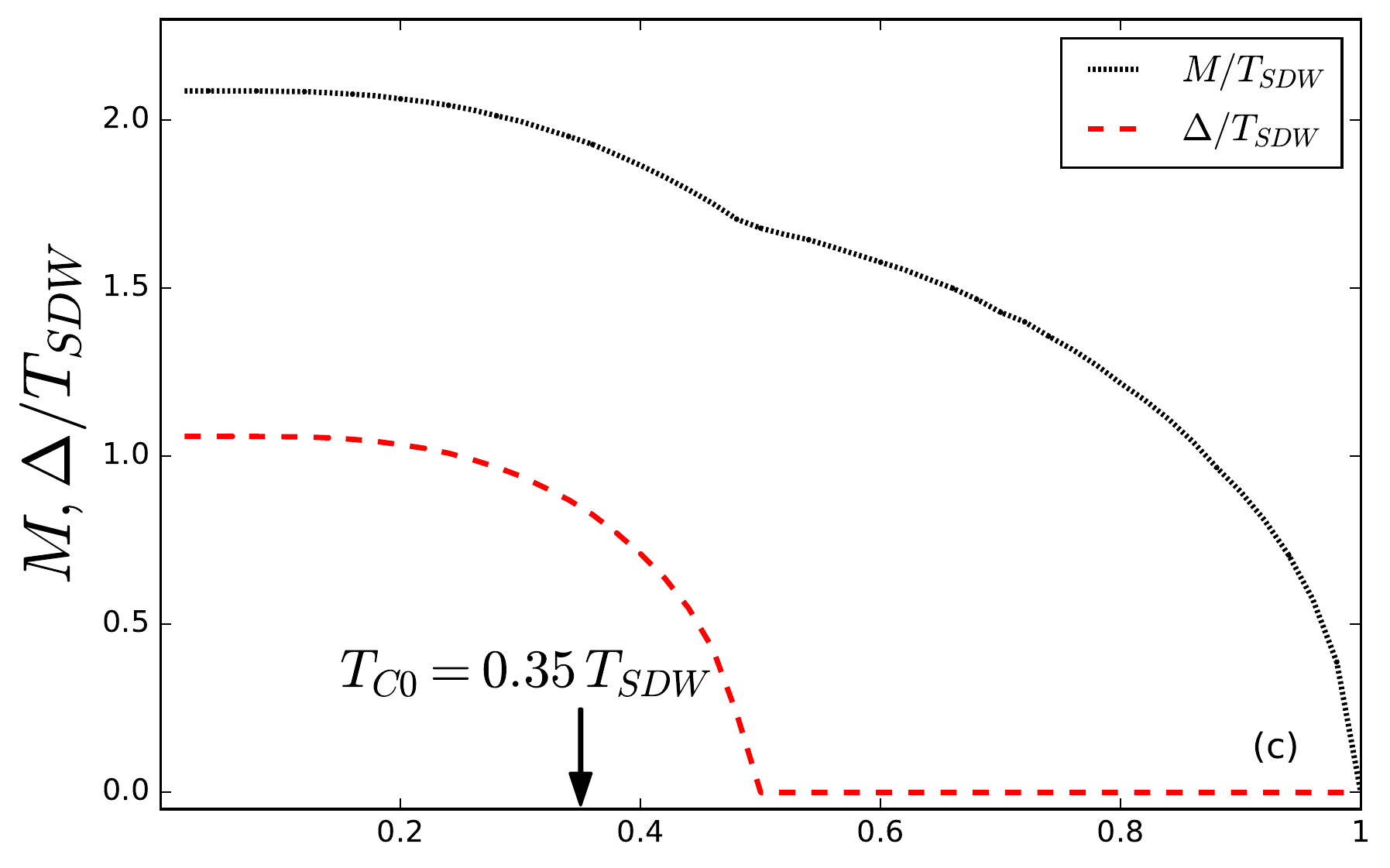}
\includegraphics[width=8cm, height=5.2cm,angle=0]{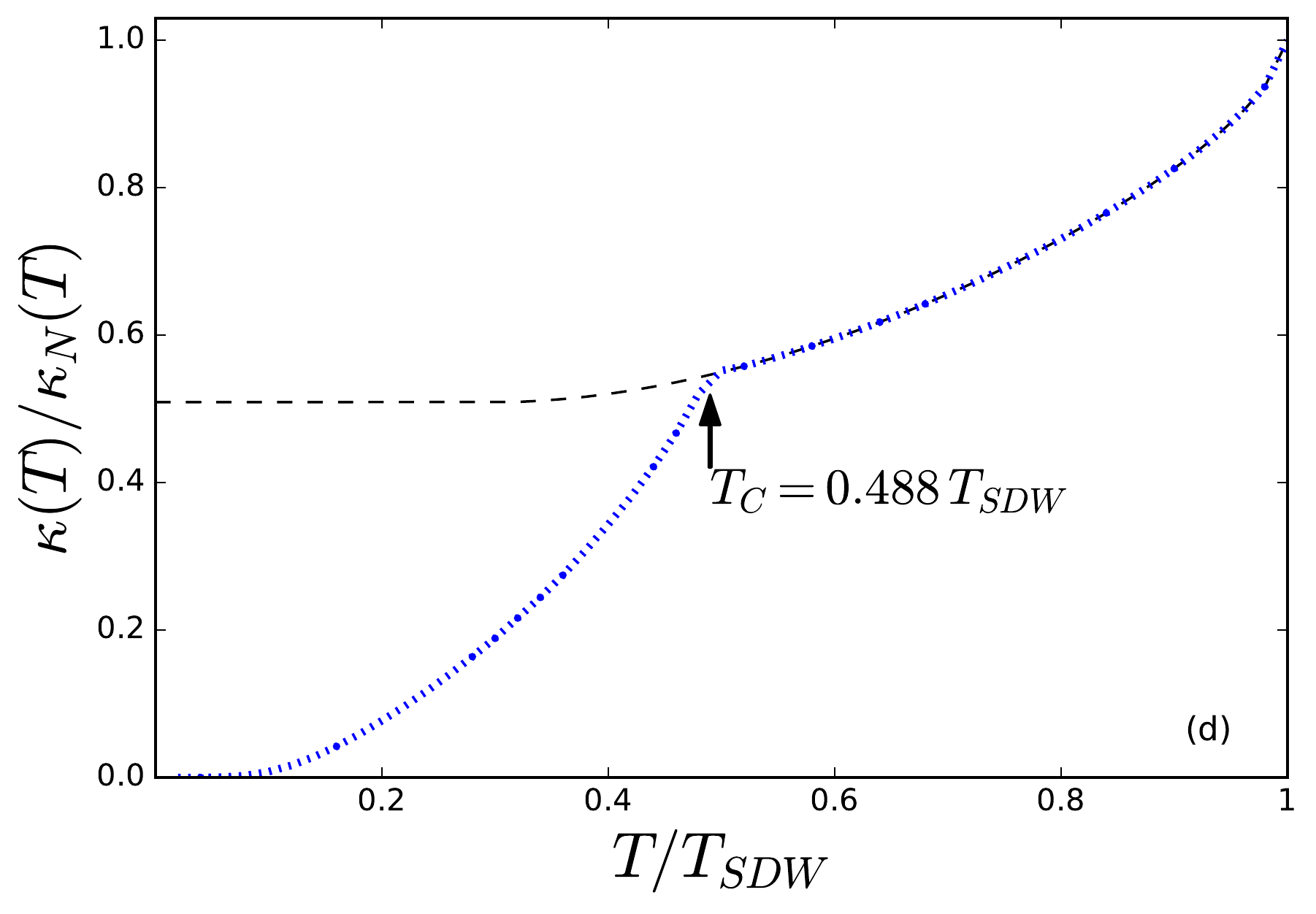}
\caption{\label{k_d22} 
Coexistence of SDW and $d_{x^2-y^2}$-SC in \textit{(E, O)} class. 
(a,b) temperature dependence of SDW and SC order parameters and thermal conductivity for $p=0.5$; 
(b,d) same for $p=0.35$. 
Inset in (b) shows FS, $d_{x^2-y^2}$ nodal lines, and the SDW-gapped regions. 
Thermal conductivity in the co-existence phase is shown by the dotted blue curve, and by the black dashed curve in a purely SDW phase} 
\end{figure}

We now turn to the discussion of the pairing states belonging to the various symmetry classes. 
{For all cases below the parameters used for FS are $t_1/2\pi T_{SDW}=100$, $t_2/2\pi T_{SDW}=10$}
 
The $d_{x^2-y^2}$ pairing state which belongs to the \textit{(E, O)} symmetry class is not competitive with the SDW, and below $T_C$ the SC order enhances the SDW order, as shown in Fig.~\ref{k_d22}. The temperature dependence of the self-consistently determined order parameters $\Delta(T)$ and $M(T)$, and of the thermal conductivity, are presented for two values of the parameter 
$$p=\frac{T_{C0}}{T_{SDW}} \;,$$ 
where $T_{C0}$ is the transition temperature of the SC order in the absence of the SDW  and $T_{SDW}$ is  the transition temperature of the SDW in the absence of the SC. 
The  $d_{x^2-y^2}$ pairing state coexists with the SDW order for all values of $p$. 
Further, the transition temperature of the SC is enhanced in the presence of the SDW. 
The onset of the SDW gaps the nested flat parts of the FS (orange shaded regions in the inset of Fig.~\ref{k_d22}(b)) leading to a weaker metallic state (remaining Fermi surface shown by cyan curves in the inset), causing the gradual fall in the thermal conductivity for $T_{C}<T<T_{SDW}$, seen in Fig.~\ref{k_d22}(b,d). 
The nodes of the $d_{x^2-y^2}$ pairing state appear under the SDW gap on nested FS parts, and thus does not result in any low-energy excitations. The sharp fall of the thermal conductivity for $T<T_{C}$, and exponential low-$T$ behavior, is characteristic of the fully gapped FS due to the simultaneous coexistence of the SDW and SC orders. 
Notice that the heat conductivity shows a kink at the co-existence transition.  

\begin{figure}
\includegraphics[width=8cm, height=5.2cm]{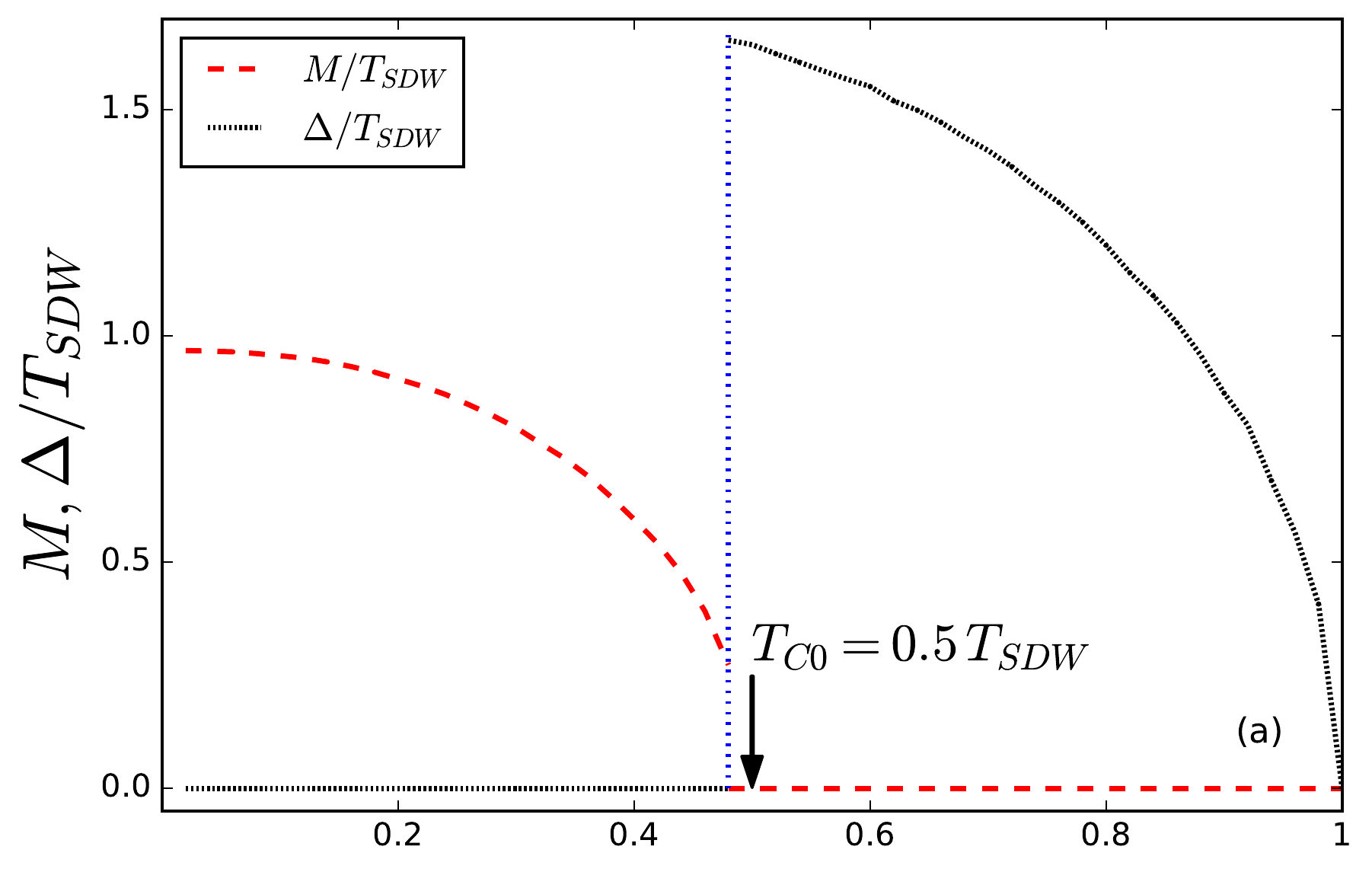}
\includegraphics[width=8cm, height=5.2cm]{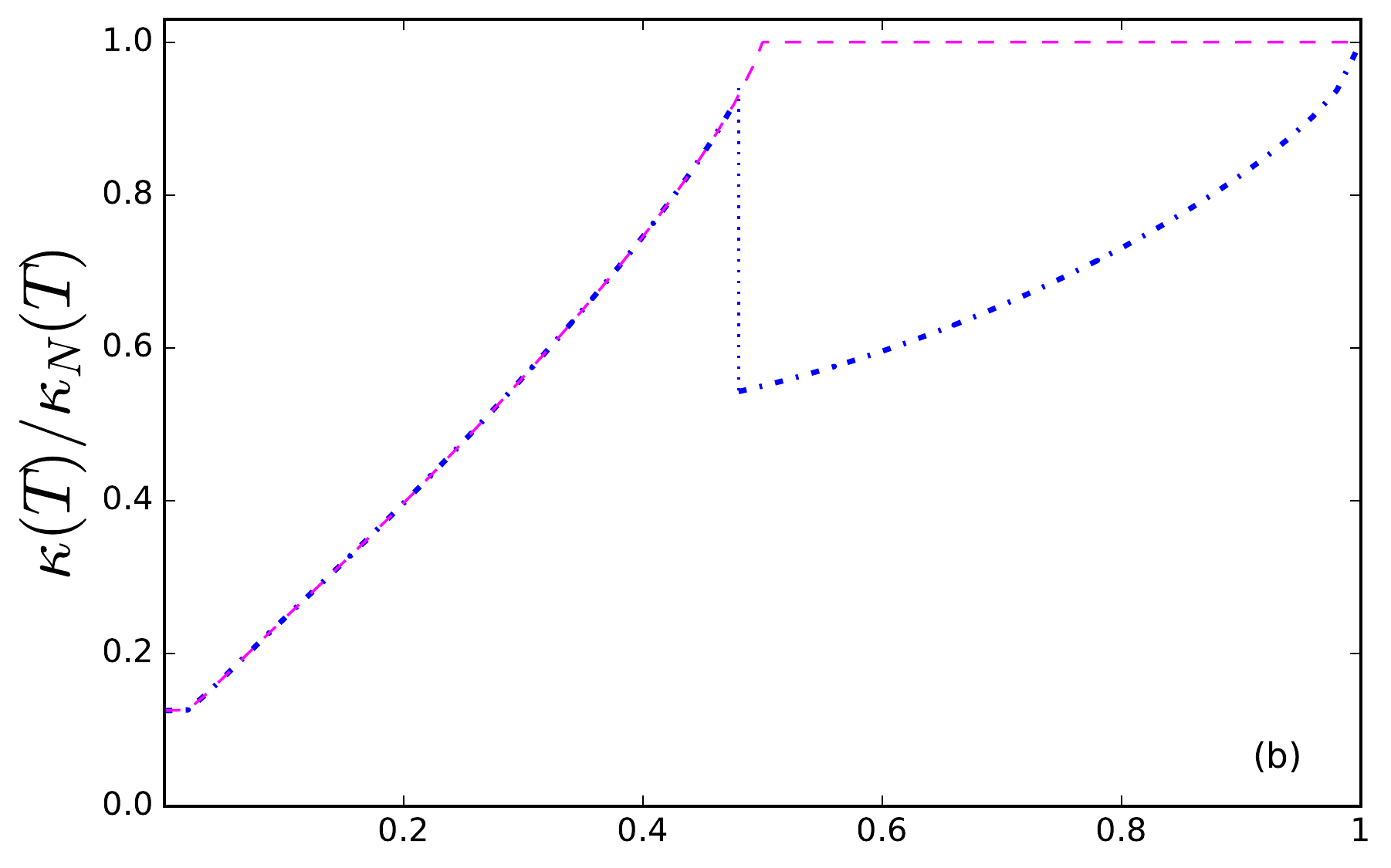}
\includegraphics[width=8cm, height=5.2cm]{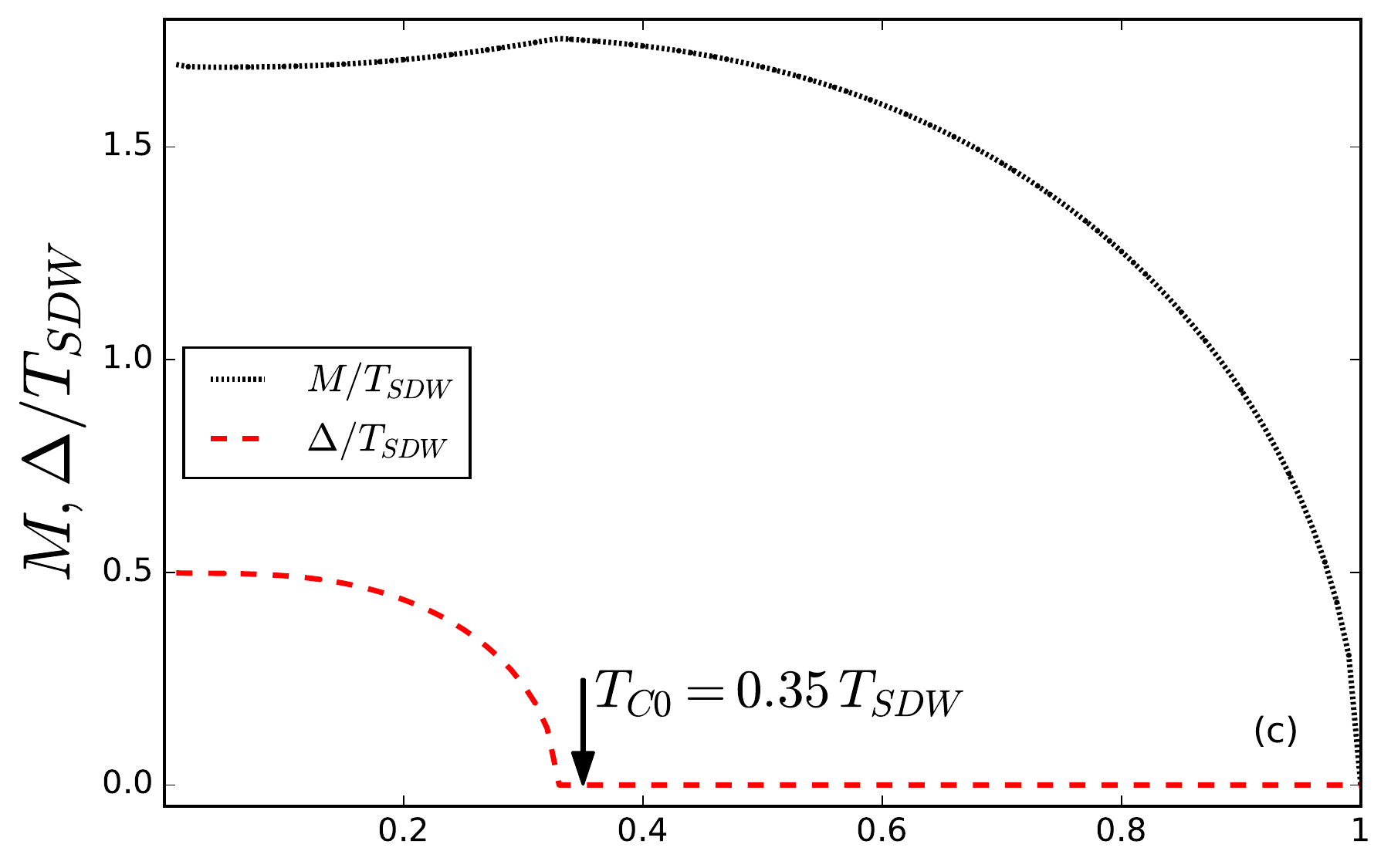}
\begin{overpic}[scale=0.46]{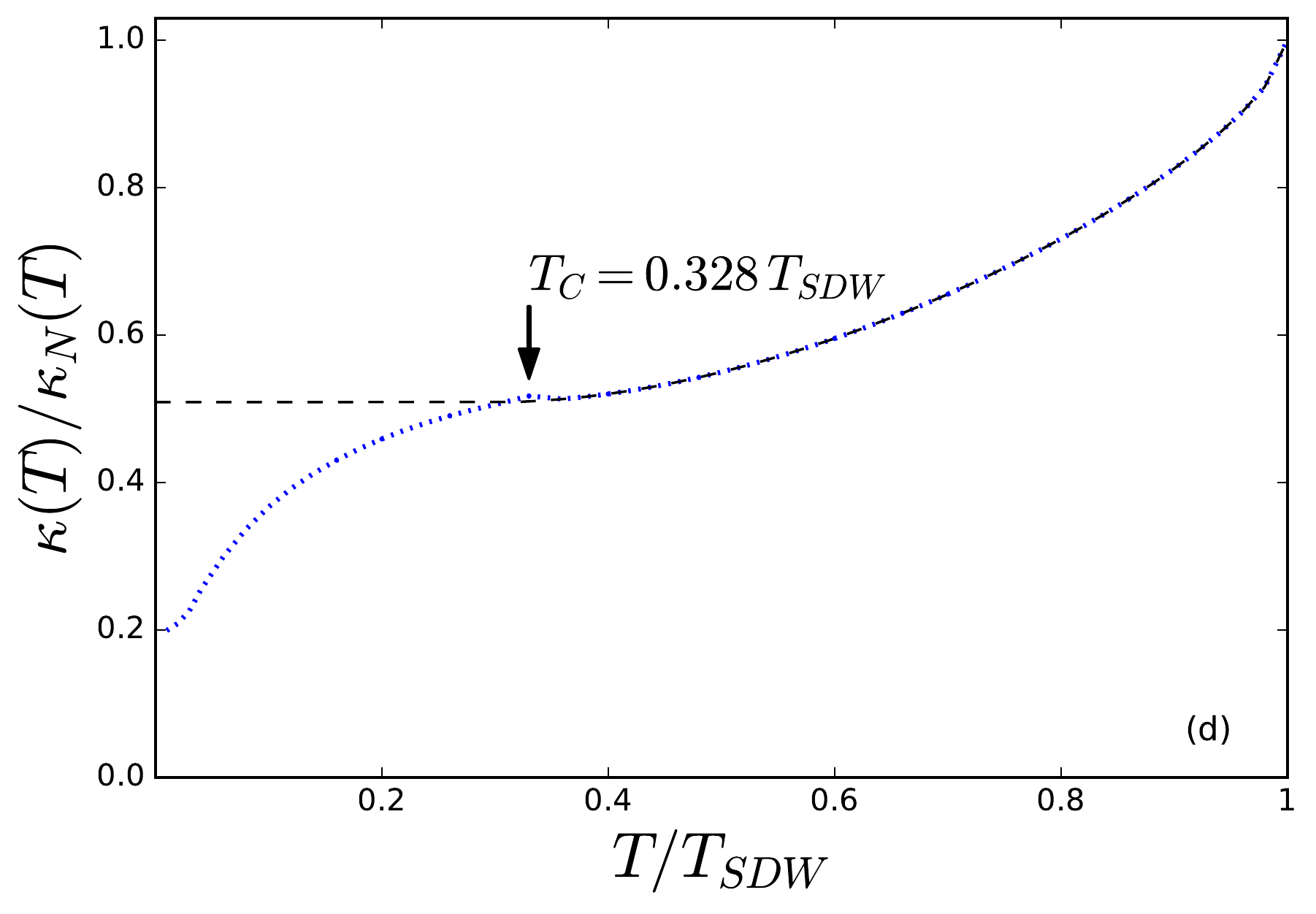}
\put(58,9.5){\includegraphics[scale=0.27]{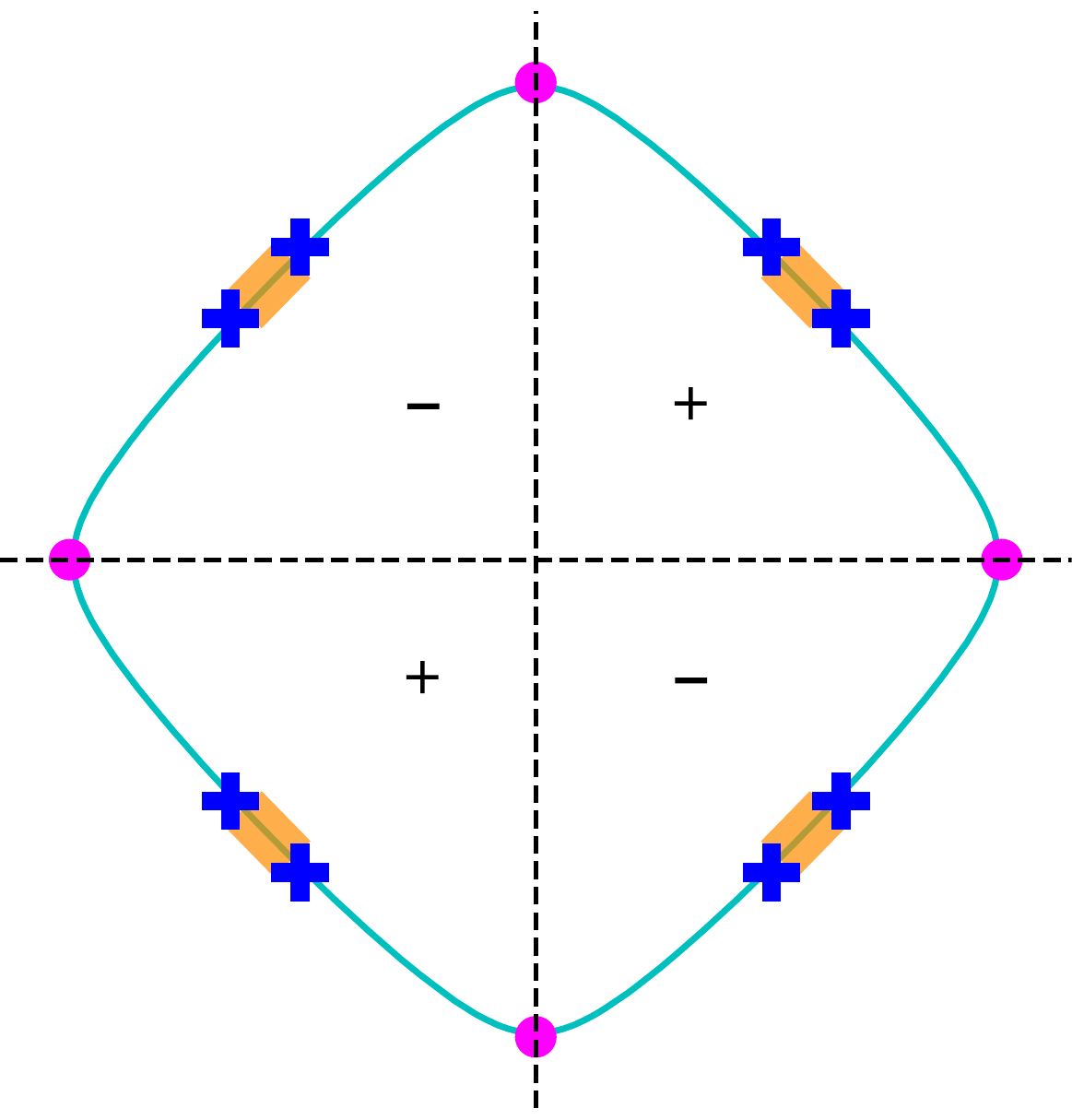}}
\end{overpic}
\caption{\label{k_dxy} Same as in Fig.~\ref{k_d22} but for interplay of SDW and $d_{xy}$-SC  in \textit{(E, E)} class. 
(a,b) for $p=0.5$ the SDW and SC states do not co-exist, switching through a 1-st order transition. Heat conductivity makes a jump reflecting $\Delta < M$ relation. 
(c,d) co-existence regime, $p=0.35$; thermal conductivity in the SDW+SC phase behaves significantly different from the pure $d_{xy}$-SC.
Inset in (d) shows the $d_{xy}$-symmetry nodes (magenta circles) and extra nodes (blue crosses) relative to the FS in $k$-space. See text for details.
}
\end{figure}

\subsection{The $(E,E)$ class}

The $d_{xy}$-wave and the isotropic $s$-wave SC pairing states belong to the \textit{(E, E)} symmetry class. 
Behavior of these states, and their signatures in thermal transport, are quite different from those for the \textit{(E, O)} symmetry class. 

We begin by discussing the $d_{xy}$ pairing state in order to contrast its behavior with the $d_{x^2-y^2}$ pairing state. 
This state does not coexist with the SDW order for all values of relative temperatures $p$. 
In  Fig.~\ref{k_dxy}(a) we show that for $p=0.5$ the SC state appears through a first order phase transition, completely replacing SDW order, whereas for $p=0.35$ it appears through a second order phase transition, and both SC and SDW order parameters are present (Fig.~\ref{k_dxy}(c)). 
However, the \textit{(E, E)} SC states compete with SDW, resulting in suppression of the magnetic order at temperatures below $T_C$, which itself is reduced. 

In the case of the first order phase transition  Fig.~\ref{k_dxy}(a), the system goes from a weak metallic phase to a purely superconducting phase. 
The SDW order $M$, that gaps only the nested parts of the Fermi surface, is replaced at $T= 0.49 T_{SDW}$ with SC gap $\Delta$, that covers more of the Fermi surface, and thus can have a lower value of the free energy, 
even at a smaller magnitude of the SC gap. 
This results in a sharp increase in the thermal conductivity. 
Behavior of the thermal conductivity for $T<T_{C}$ in this case is the same as that of the $d_{xy}$ pairing state in the absence of the SDW. The dashed red curve is the appropriately scaled thermal conductivity in the pure SC state, from Fig.~\ref{k_sc}. 

When the SC and SDW order can coexist, e.g. for the case of $p=0.35$ shown in Fig.~\ref{k_dxy}(c,d), behavior of thermal transport is very unusual. Below $T_{SDW}$, a part of the FS gets gapped with $M$, indicated by the shaded orange in the sketch in inset of \ref{k_dxy}(d). $\kappa(T)$ drops, but gets saturated at a finite value due to the remaining FS, shown by the cyan lines, which gives a weaker-than-normal metallic state that we denote as SDW-metallic state. One expects that at the onset of $d_{xy}$ order with symmetry nodes (magenta dots) on this FS, the heat conductivity would show a somewhat similar behavior to the one for the pure SC state, as in Fig.~\ref{k_sc}. 
%
This is indeed the case, as can be seen in \ref{k_dxy}(d). The only quantitative difference is due to 
appearance of the extra non-symmetry nodes in the SDW+SC state, discussed in section \ref{sec:model}.\ref{sec:diagH}. They arise near the SDW-gapped region, and marked as the blue crosses in the sketch.
Since the excitation gap collapses in both symmetry node and the extra node, the SC order parameter $\Delta_\vk$, although growing in amplitude below $T_C$, does not efficiently gap the FS between the nodes, resulting in a more gradual reduction of ${\kappa(T)}/{\kappa_N(T)}$ below $T_{C}$. In the low temperature limit the extra nodes result in relative enhancement of the residual thermal conductivity.   

\begin{figure}
\centering
\includegraphics[width=8cm, height=5.2cm]{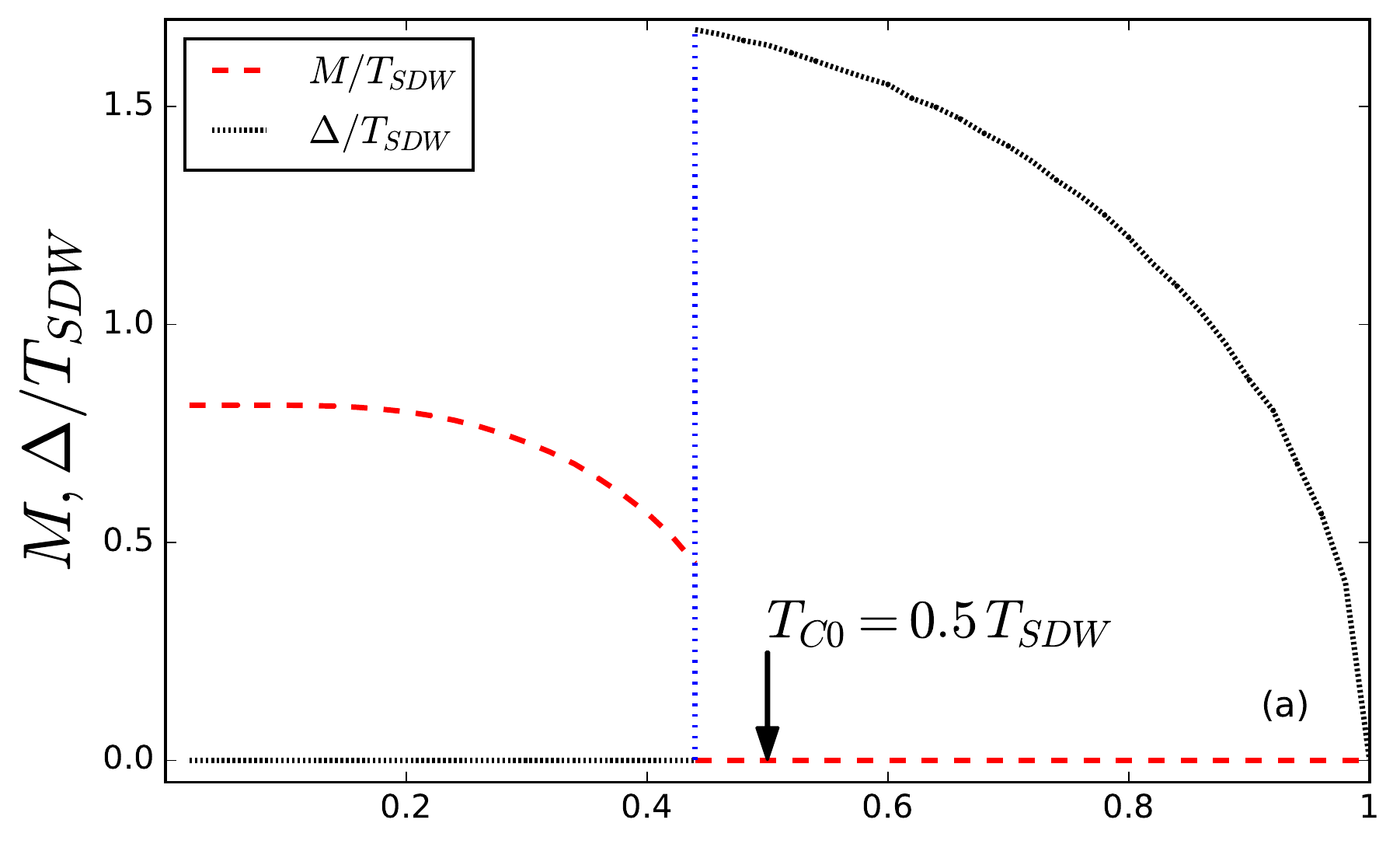}
\includegraphics[width=8cm, height=5.2cm]{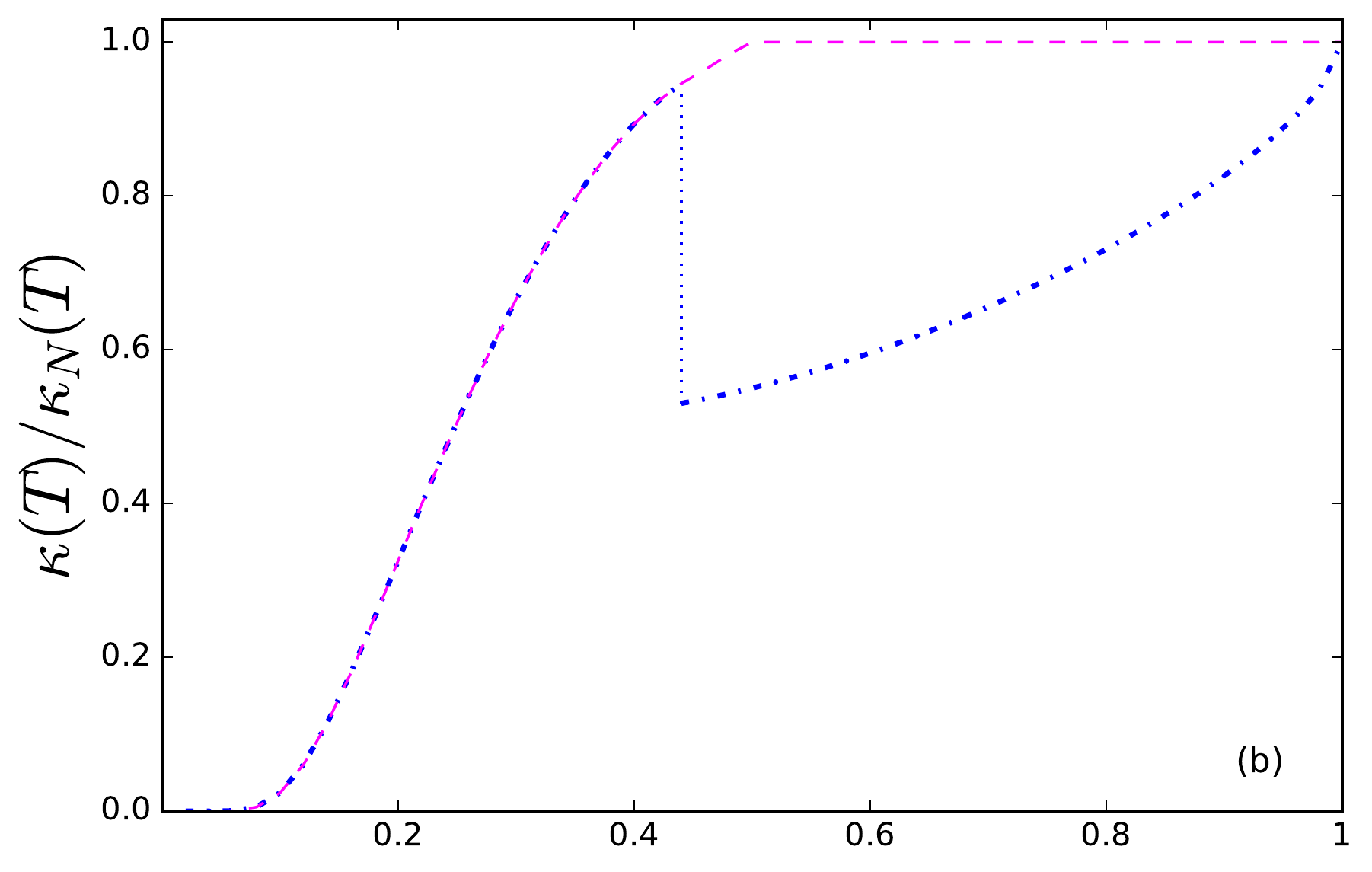}
\includegraphics[width=8cm, height=5.2cm]{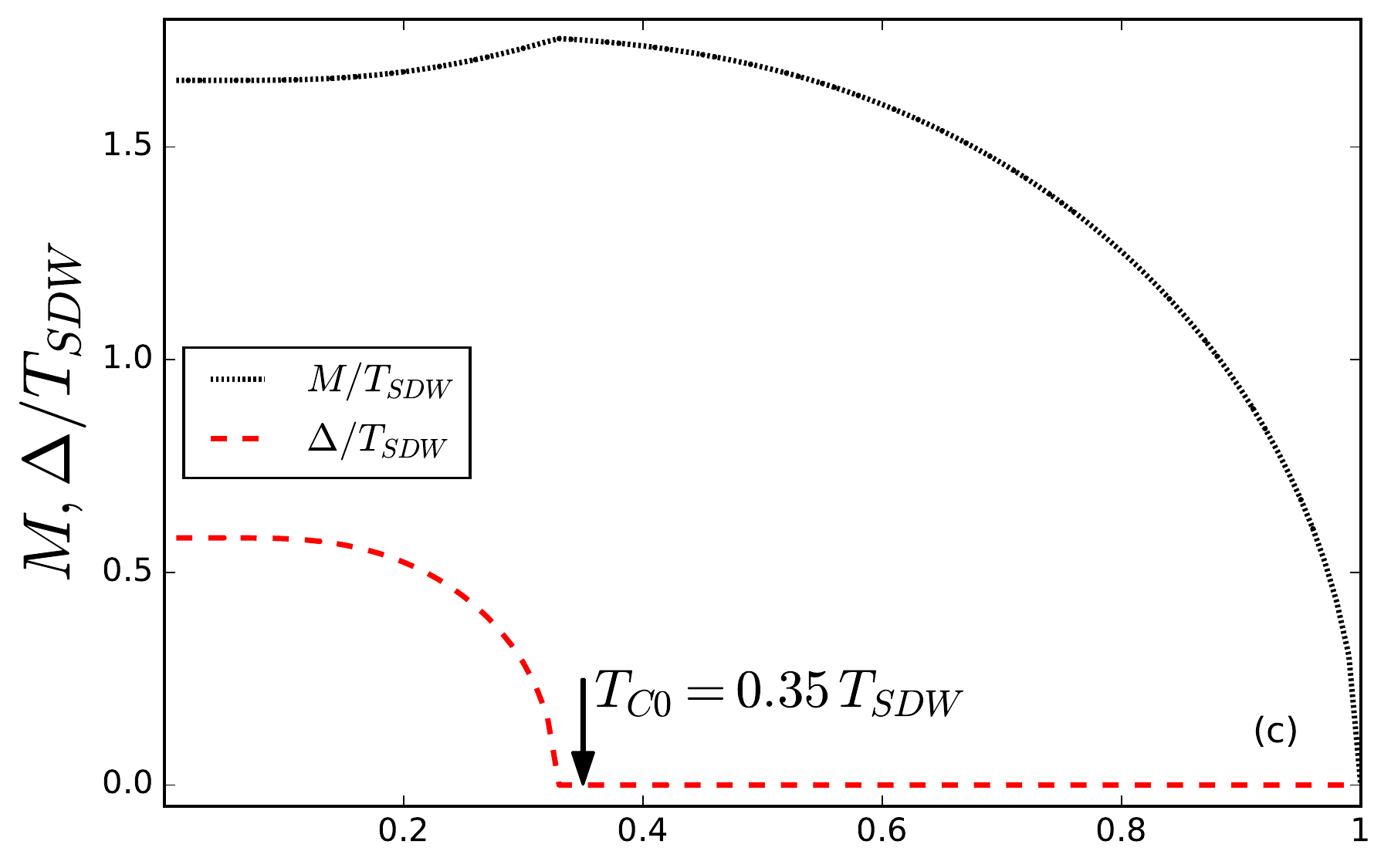}
\begin{overpic}[scale=0.44]{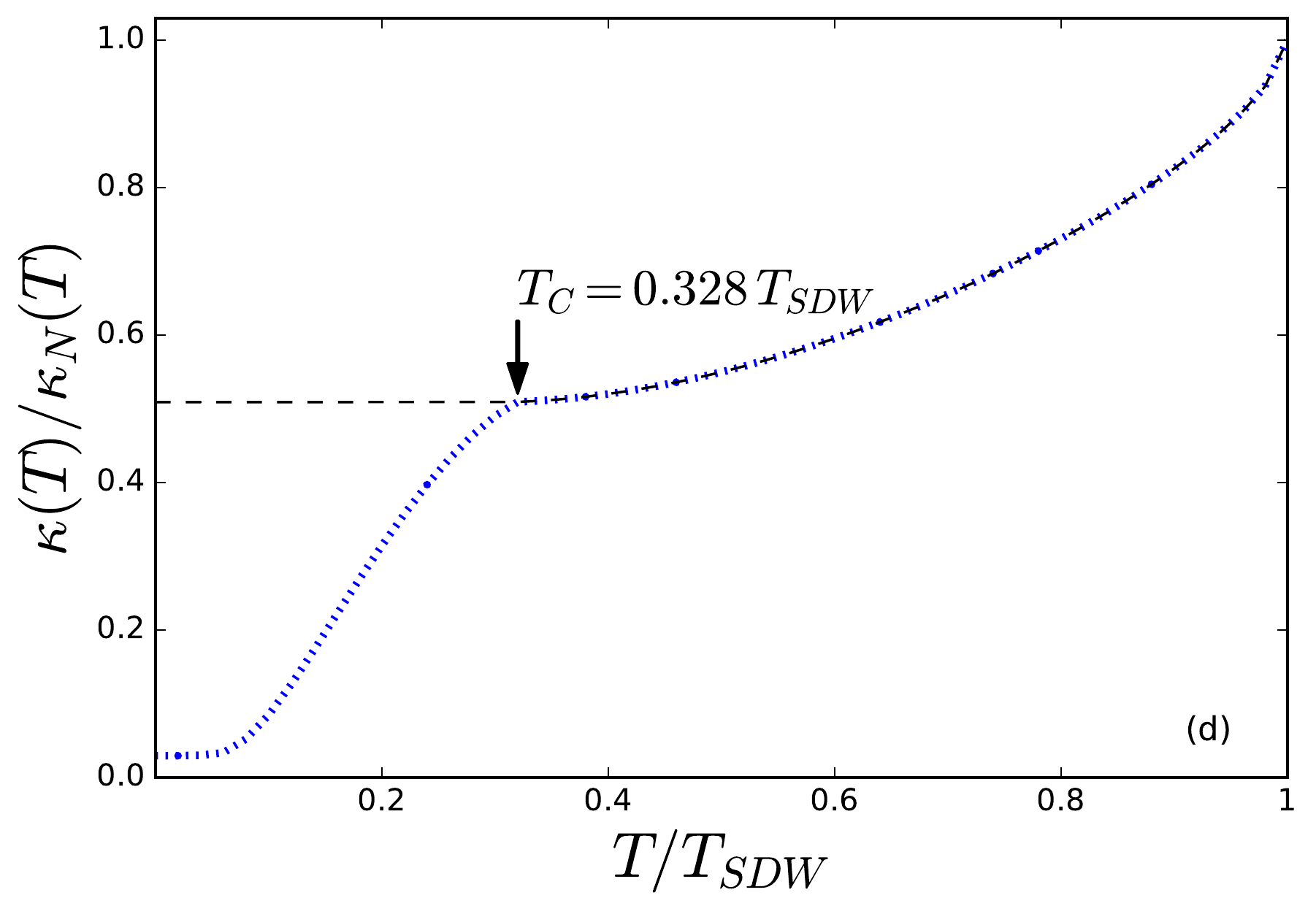}
\put(58,9){\includegraphics[scale=0.27]{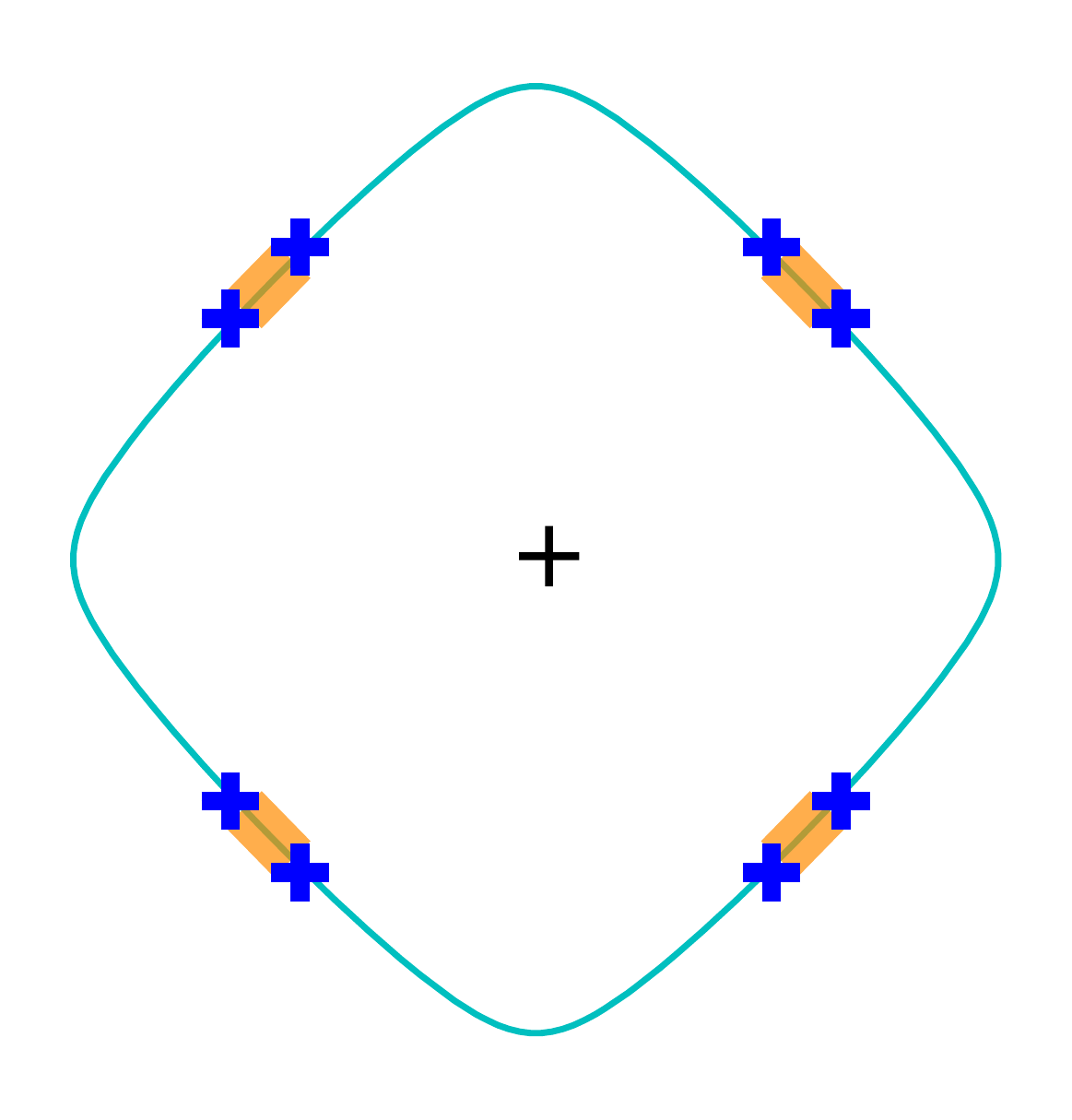}}
\end{overpic}
\caption{\label{k_s} Interplay of SDW and $s$-SC 
(a,b) temperature dependence of $M,\Delta$, and thermal conductivity when $T_{C0}/T_{SDW}=0.5$. 
The SDW and SC states do not co-exist, similarly to $d_{xy}$ case.  
(c,d) Co-existence is possible for lower $T_{C0}/T_{SDW}=0.35$; 
{thermal conductivity at low temperatures reaches a finite value due to the emergent nodes}, shown by the blue crosses in the inset of (d). See text for details.
}

\end{figure}

Fully gapped $s$-wave state shows similar coexistence pattern: it fully replaces SDW order for strong SC pairing, resulting in a sudden jump of physical observables, Fig.~\ref{k_s}(a,b).
For $p=0.35$, the $s$-wave SC state appears through a second order phase transition, resulting in SDW+SC co-existence, Fig.~\ref{k_s}(c). 
At $T_C$ the emerging SC order gaps the SDW-metallic state resulting in a suppression of $\kappa$.
Again, the gapless excitations in the additional nodes (blue cross marks in the inset of Fig.~\ref{k_s}(d)), at low temperatures result in finite thermal conductivity, eliminating the exponential character of the fully-gapped $s$-wave heat transport. 

\section{Conclusion} 
\label{sec:conclusion}

We have considered a single-band electronic system where spin-singlet Superconducting order can appear inside a collinear Spin-density-wave phase, at the mean-field level. 
It is based on a tight-binding model on a square lattice with a commensurate SDW with ordering vector $\mathbf{Q}=(\pi,\pi)$.
Coexistence of the SC and SDW orders is controlled by selecting a band structure with a Fermi surface, such that only a part of it is nested supporting SDW order, leaving the other part for SC.  
The amplitudes of the SC and SDW orders were determined self-consistently at all temperatures.  

The nature of the coexisting phase depends, most importantly, on the properties of the SC order parameter connected by the nesting vector $\vQ$ 
If the SDW order mixes up pairs with $\Delta_{\vk +\vQ}=-\Delta_{\vk}$, as is the case for the $d_{x^2-y^2}$ SC symmetry, the two orders attract each other and naturally coexist\cite{machida3}.
Mixing states with $\Delta_{\vk +\vQ}=\Delta_{\vk}$ ($d_{xy}$- or $s$-wave) results in competition of SDW and SC, although they can still coexist for weak enough SC state arising inside the SDW phase.

One of the most interesting differences between the two versions of SC+SDW mixture is the spectrum of low-energy excitations. For SDW+$d_{x^2-y^2}$ the nodes of the SC order appear on the nested parts of the FS and thus appear under the SDW gap, resulting in the fully gapped system. On the other hand, in SDW+$d_{xy}$ the symmetry-protected SC nodes appear on the non-nested part the Fermi surface. In addition to those, we found additional set of robust nodes, appearing on the boundary of the folded Brillouin zone. These nodes are the remnants of the SDW-state Fermi surface, and exist even in the $s$-wave superconducting state. 
They form an anisotropic Dirac cone of low-energy excitations.

Temperature dependence of the electronic heat conductivity in the SDW+SC system was computed using Boltzmann transport equation method, where the impurity scattering collision integral and quasiparticle lifetime were determined (in Born limit) from the correct coherence factors of the co-existence phase. 
Our numerical analysis shows that there are significant differences in the thermal conductivity behavior that are determined by the symmetry of the order parameter, FS topology, and the nodal structure of the co-existence phase. 

For the SDW+$d_{x^2-y^2}$ combination, the nodal structure of SC order parameter is immersed under the SDW gap producing only gapped excitations that result in the rapid drop of the thermal conductivity below the second-order co-existence transition, and typical exponentially-small residual $\kappa(T)/T$. 

On the other hand, in SDW+$s$,$d_{xy}$ system, the two orders may completely avoid each other, resulting in the trivial first-order jump in heat conductivity. However, the most interesting situation arises when SC does not replace SDW completely at low temperature, and they co-exist. The nodal quasiparticles are preserved in this case, and even new Dirac-like excitations appear in both $d_{xy}$ and $s$-wave systems. {These low-energy excitations lead to a finite residual $\kappa/T$ in the $T\to0$ limit for both the SDW+$s$,$d_{xy}$ systems.} 


\section*{Acknowledgements}
Numerical work was done on the Pacific Research Platform's Nautilus HyperCluster. S.S.C would like to thank Nazmul Kazi for 
help with implementation of the numerical analysis. 

\appendix

\section{2-step diagonalization}
\label{appA}
We wish to clarify certain aspects of the diagonalization procedure that we have employed in this paper and compare it with  previous work done on similar models by several authors \cite{Ismer}. We have diagonalized the full mean field Hamiltonian $\hat{\cH}(\vk )$ in (\ref{ham matrix}) using a unitary  Bogoliubov transformation $\hat{B}(\mathbf{k})$ by numerically computing eigenvectors in the RBZ. In literature a  `two-step' procedure is often employed to diagonalize the model Hamiltonian (\ref{ham matrix}),  which yields  identical results to our case, provided all pairing terms are properly accounted for. Step one of the two-step process involves diagonalizing the first two terms in (\ref{ham full}) via a unitary transformation by introducing new quasiparticle operators $\alpha_{\vk }, \beta_{\vk }$ for the two SDW bands with dispersions $E_{\vk }^{\alpha,\beta} =\xi_{\vk }^{+} \pm \sqrt{(\xi_{\vk }^{-})^2 + M^2 }$. Namely the first two terms in the Hamiltonian are written as 
\begin{eqnarray}
\begin{aligned}
&H_0 + H_{SDW}=\sum\limits_{\sigma}\sum\limits_{\vk \in RBZ} \psi^{\dagger}_{\vk i}h_{1\vk ij}\psi_{\vk j}    \\
&h_{1\vk ij} =
\begin{pmatrix}

\xi_{\vk } & \sgn(\sigma) M \\

 \sgn(\sigma) M & \xi_{\mathbf{k + Q}} \\
\end{pmatrix}
\label{SDW_ham matrix}
\end{aligned} 
\end{eqnarray}  
where  $\psi^{\dagger}_{\vk i}=(c_{\vk \sigma}^{\dagger}, c_{\mathbf{k+Q}\sigma}^{\dagger})$ defines the Nambu basis. The above hamiltonian is then diagonalized using the following Bogoliubov transformation 
\begin{align}
\begin{split}
&c_{\vk \sigma}=  u_{\vk }\alpha_{\vk \sigma} - \sgn(\sigma)v_{\vk }\beta_{\vk \sigma}\\
&c_{\mathbf{k+Q}\sigma}= \sgn(\sigma)v_{\vk }\alpha_{\vk \sigma} + u_{\vk }\beta_{\vk \sigma}\\
\end{split}
\label{SDW_BT}
\end{align}
where $u_{\vk }= \sqrt{\frac{1}{2}(1+\frac{\xi^-_{k}}{\Pi_{\vk }})}$,  $v_{\vk }= \sqrt{\frac{1}{2}(1-\frac{\xi^-_{k}}{\Pi_{\vk }})}$ with $\Pi_{\vk }= \sqrt{(\xi_{\vk }^{-})^2 + M^2}$.  The diagonalization reduces (\ref{SDW_ham matrix}) to 
\begin{align}
H_0 +H_{SDW}= \sum\limits_{\substack{\vk \in RBZ \\ \sigma}} E^{\alpha}_{\vk }\alpha^{\dagger}_{\vk \sigma}\alpha_{\vk \sigma}+E^{\beta}_{\vk }\beta^{\dagger}_{\vk \sigma}\beta_{\vk \sigma}
\label{SDW_diag}
\end{align}
In step two, the same  unitary transformation (\ref{SDW_BT}), is applied to the superconducting term $H_{SC}$ in (\ref{ham full}), which when combined with (\ref{SDW_diag}), results in the following mean field Hamiltonian  

\begin{eqnarray}
\begin{aligned}
&H_0 +H_{SDW} + H_{SC}=\frac{1}{2}\sum\limits_{\vk \in RBZ} \gamma^{\dagger}_{\vk i}h_{\vk ij}\gamma_{\vk j}    \\
&h_{\vk ij} =
\begin{pmatrix}

E_{\vk }^{\alpha} & \Delta_{\vk }^{\alpha}  & 0 &\Delta_{\vk }^{\alpha \beta}  \\

\Delta_{\vk }^{\alpha} & -E_{\vk }^{\alpha} & -\Delta_{\vk }^{\alpha \beta} & 0 \\

0 &-\Delta_{\vk } ^{\alpha \beta} & E_{\vk }^{\beta} & \Delta_{\vk }^{\beta} \\

\Delta_{\vk }^{\alpha \beta} & 0 & \Delta_{\vk }^{\beta}  & -E_{\vk }^{\beta} \\

\end{pmatrix}
\label{2step ham matrix}
\end{aligned} 
\end{eqnarray}
where $\gamma^{\dagger}_{\vk i}=(\alpha_{\vk \uparrow}^{\dagger}, \alpha_{-\vk \downarrow},\beta_{\vk \uparrow}^{\dagger},\beta_{-\vk \downarrow})$ defines  the Nambu basis. The superconducting order parameters dressed by the SDW coherence factors  are given by $\Delta_{\vk }^{\alpha}= u^2_{\vk }\Delta_{\vk }- v^2_{\vk }\Delta_{\mathbf{k+Q}}$ , $\Delta_{\vk }^{\beta}= u^2_{\vk }\Delta_{\mathbf{k+Q}}- v^2_{\vk }\Delta_{\vk }$ and $\Delta_{\vk }^{\alpha \beta}= u_{\vk }v_{\vk }(\Delta_{\vk }+\Delta_{\mathbf{k+Q}} )$.  If one neglects the off-diagonal blocks in the above Hamiltoninan i.e. inter-band pairing terms of the form $\langle \alpha_{\vk \uparrow}^{\dagger} \beta^{\dagger}_{-\vk \downarrow}\rangle $ etc. , then (\ref{2step ham matrix}) can be diagonalized by two independent Bogoliubov transformations which yield the energy dispersions \cite{Ismer} $\mathcal{E}^{\gamma}_{\vk }= \sqrt{(E^{\gamma}_{\vk })^2  + (\Delta_{\vk }^{\gamma})^2}$ where $\gamma=(\alpha,\beta)$. We do not neglect the inter-band pairing terms of the form $\langle \alpha_{\vk \uparrow}^{\dagger} \beta^{\dagger}_{-\vk \downarrow}\rangle $ when diagonalizing  (\ref{2step ham matrix}). If we diagonalize (\ref{2step ham matrix}) keeping the off diagonal terms\cite{romer}  , we get the following dispersion relation 

\begin{align}
\begin{split}
E^{2}_{1,2}=&\frac{1}{2}\bigg(\Xi_{\vk } \pm \Sigma_{\vk }\bigg) \\
\Xi_{\vk }=&\bigg[(E_{\vk }^{\alpha})^2 + (E_{\vk }^{\beta})^2 + (\Delta_{\vk }^{\alpha})^2 + (\Delta_{\vk }^{\beta})^2 + 2\Delta_{\vk }^{\alpha \beta}\bigg]\\
\Sigma_{\vk }=&\bigg[\Pi_{\vk }^2 -4\bigg((E_{\vk }^{\beta})^2(\Delta_{\vk }^{\alpha})^2 + 2E_{\vk }^{\alpha}E_{\vk }^{\beta}(\Delta_{\vk }^{\alpha \beta})^2\\ 
+& \big((\Delta_{\vk }^{\alpha \beta})^2 + \Delta_{\vk }^{\alpha}\Delta_{\vk }^{\beta}\big)^2 + (E_{\vk }^{\alpha})^2\big((E_{\vk }^{\beta})^2 + (\Delta_{\vk }^{\beta})^2\big)
\bigg)\bigg]^{\frac{1}{2}}
\end{split}
\label{2 step eigenvalues}
\end{align}
Upon substituting the expressions for $\Delta_{\vk }^{\alpha}, \Delta_{\vk }^{\beta},\Delta_{\vk }^{\alpha \beta}, E_{\vk }^{\alpha}$ and  $E_{\vk }^{\beta}$, one recovers the eigenvalues given in (\ref{eigen}). 
\vspace{1.5cm}

\bibliography{ref}
\bibliographystyle{apsrev.bst}

\end{document}